\documentclass[aps,twocolumn,pra,superscriptaddress,showpacs,tightenlines]{revtex4-1}
\usepackage{amsmath}
\usepackage{graphicx}
\usepackage{epsfig}
\usepackage{amsfonts}
\usepackage{color}
\usepackage{epstopdf}
\usepackage{hyperref}
\usepackage[T1]{fontenc}
\usepackage[latin9]{inputenc}
\usepackage{amsmath}
\usepackage{amssymb}
\usepackage{graphicx}
\usepackage{esint}
\usepackage{footnote}

\begin{document}

\newcommand{\nn}{\nonumber}
\newcommand{\ms}[1]{\mbox{\scriptsize #1}}
\newcommand{\msi}[1]{\mbox{\scriptsize\textit{#1}}}
\newcommand{\dg}{^\dagger}
\newcommand{\smallfrac}[2]{\mbox{$\frac{#1}{#2}$}}
\newcommand{\Tr}{\text{Tr}}
\newcommand{\ket}[1]{|#1\rangle}
\newcommand{\bra}[1]{\langle#1|}

\newcommand{\SMstart}{\begingroup\color{black}}
\newcommand{\SMend}{\endgroup}
\newcommand{\pfpx}[2]{\frac{\partial #1}{\partial #2}}
\newcommand{\dfdx}[2]{\frac{d #1}{d #2}}
\newcommand{\half}{\smallfrac{1}{2}}
\newcommand{\s}{{\mathcal S}}
\newcommand{\jord}{}
\newcommand{\kurt}{}

\makeatletter


\newcommand*\LyXZeroWidthSpace{\hspace{0pt}}
\providecommand{\tabularnewline}{\\}

\title{{\color{black}Unified Coherent Error Correction Code: A Unified Framework for Combating Decoherence in Universal Quantum Computing with Identical Particles}}

\author{S. L. Wu }
\affiliation{School of Physics and Materials Engineering,
Dalian Nationalities University, Dalian 116600 China}
\affiliation{Department of Physics, University of the Basque Country UPV/EHU, 48080 Bilbao, Spain}



\author{Lian-Ao Wu }
\email{lianao.wu@ehu.es}
\affiliation{Department of Physics, University of the Basque Country UPV/EHU, 48080 Bilbao, Spain}
\affiliation{IKERBASQUE Basque Foundation for Science, 48013 Bilbao, Spain}
\affiliation{EHU Quantum Center, University of the Basque Country UPV/EHU, Leioa, Biscay 48940, Spain}

\date{\today}

\begin{abstract}

{We present a universal quantum computing architecture based on identical particle qubits (IPQs), where we find that the first-order IPQ - bath interaction fundamentally differs from the conventional first-order qubit-bath interaction. This key distinction necessitates a redesign of existing strategies to fight decoherence. We propose that the simplest quantum error correction code can be realized directly with the {\color{black}physically implementable} qubit, provided that conventional correction and restoration are generalized beyond unitary operations to employ physically implementable reversal operations - naturally {\color{black}placing a physically implementable qubit and a quantum error correction code against collective decoherence between two bosonic modes on equal footing}. We further demonstrate that dynamical decoupling (DD) remains effective within this unified framework, and that a decoherence-free subspace (DFS) -like structure emerges. Unlike previous approximate treatments, our analytically solvable IPQ-Bath model enables rigorous testing of these strategies, with numerical simulations validating their effectiveness.}


\end{abstract}

\maketitle

\section{Introduction}
Quantum decoherence remains the primary obstacle to realizing practical quantum computers, making error correction and suppression
 essential challenges
in their development~\cite{Schlosshauer2019}.
Several strategies to fight decoherence, including the closed-loop QEC, open-loop DD and DFS, have been proposed and implemented based on first-order qubit-bath interactions across various physical platforms~\cite{Ripoll2003,Cao2023,Arunkumar2023,Jain2024,Leon2021}. 
Closed-loop QEC theories and practices aim to protect quantum information by encoding logical qubits into larger Hilbert spaces including multi-qubit space and bosonic space, such as the surface code, color codes, concatenated codes, and bosonic binomial codes to enable continuous-variable systems for hardware-efficient protection against photon loss~\cite{PRX}. In recent years, experimental progress has accelerated significantly: repeated error correction using stabilizer codes has demonstrated logical error rates well below physical rates on trapped-ion platforms~\cite{Paetznick2021}, while bosonic codes, including cat and binomial codes, have been successfully implemented in superconducting cavities with concatenated architectures ~\cite{Preskill2025}, and error correction with high-dimensional bosonic binormial code~\cite{Binormial2023} and GKP codes~\cite{Brock2025}. These efforts highlight both the diversity of QEC strategies and the growing maturity of fault-tolerant operations across platforms. \textcolor{black}{Although extending the lifetimes of encoded logical qubits beyond those of the best available physical qubits remains theoretically elusive, experimental demonstrations have been reported on a case-by-case basis, such as in Ref.~\cite{Binormial2023}}.

In this work, we investigate a general scheme that encodes qubits using identical particles,
where quantum information is stored in two quantum states or levels, each occupied by one of two identical particles. This identical
particle qubit scheme leverages intrinsic degrees of freedom, similar to encoding in cold atoms \cite{Ripoll2003, Teoh2023,Chou2024,Grassl1997} or dual-rail representations in  linear optical quantum computing \cite{Knill2001,Shapiro2006,Li2015, Bartolucci2023,Chou2024}. Such a scheme has demonstrated favourable performance against errors and improved robustness against Pauli errors and applications
\cite{Berdou2023, Reinhold2020, Adam2014,Wu2009}. { Unusually,  we find that the first-order system-bath (IPQ-B) interaction in IPQ setting differs fundamentally from the conventional first-order qubit-bath interaction assumed in standard error correction schemes. As a result, conventional approaches to fight decoherence and noise, including QECCs, must be redesigned or reformulated to account for the structure of the first-order IPQ-B interaction. }

Our analysis reveals that the simplest QECC is naturally realized by the IPQ itself, where system errors can be detected { by measuring the parity of the IPQ \cite{Lin2020, Sun2014,Proctor2017}. However, as a code, the IPQ does not satisfy the standard QEC criteria when errors arise from the first-order IPQ-B interaction. Rather than relying on a unitary recovery operation, here we propose using a physically realizable reversal process implemented through a joint unitary operation on the IPQ and an ancillary qubit, followed by measurement of the ancilla. }
Furthermore, this first-order IPQ-B interaction leads to analytically solvable dynamics and reveals exact symmetries even when IPQs are embedded in an environment.  We identify a DFS-like and associated gate operations that are inherently robust against such interactions~\cite{Friesen2017}.

Dynamical decoupling is effective for IPQs: leakage-elimination operators (LEOs) commute with--and are independent of--the universal gate set, so control adds no extra errors \cite{Wu2002,Jing2015}; DD performance is strong in practice \cite{Ezzell2023}. Under suitable conditions this framework can eliminate quantum noise, enabling a unified code that integrates QECC, DD, and DFS for a robust, scalable route to fault-tolerant quantum computing with identical particles.

In this Article, we introduce an identical-particle qubit (IPQ) encoding that stores one logical qubit in two modes occupied by a single identical particle. We first present the encoding and its controllable generators, then analyze its error-correction properties and compare it with representative bosonic encodings. We next study the system--bath dynamics in closed form, identify how collective and individual reservoirs affect both the encoded subspace and leakage, and show how leakage-elimination operators (LEOs) suppress the dominant error channels. Finally, we discuss a two-qubit gate implementation and quantify its performance in the presence of finite-temperature noise.

The paper is organized as follows. Section~\ref{sec:IPQ} introduces the encoding, its generators, and its error-correction features, together with comparisons and performance benchmarks. Section~\ref{sec:exact} presents the main dynamical results for the IPQ coupled to bosonic environments and the resulting control protocols. Section~\ref{sec:indiv} analyzes individual reservoirs and presents numerical results. Technical derivations are collected in the Appendices.
\section{Identical-particle qubits as a self-protected error-correcting encoding}\label{sec:IPQ}

In this section we define the IPQ encoding and the associated controllable generators, and then characterize its error-correction properties in a form parallel to standard qubit QECCs. We also compare the IPQ with representative bosonic encodings to clarify when identical-particle hardware offers an advantage.

\subsection{Encoding and generators}
We consider a system comprising two modes of identical particles with creation operators
\( {c}_{0}^{\dagger}\) and \( {c}_{1}^{\dagger}\). The information is encoded in
the two states as a qubit, specifically an identical particle qubit (IPQ) spanned by
$|0\rangle =  {c}_{0}^{\dagger}|\text{vacuum}\rangle$ and $|1\rangle =  {c}_{1}^{\dagger}|\text{vacuum}\rangle
$, where \(|\text{vacuum}\rangle\) denotes the vacuum state.
\textcolor{black}{
Here \(c_0^\dagger\) and \(c_1^\dagger\) create two internal states, modes,
or basis states of the same physical qubit, atom, or site, with the total
particle number fixed to one. Thus, the IPQ is a physically implementable
two-internal-state qubit rather than a logical qubit encoded in two separate
physical qubits. Although this representation is mathematically analogous to
the bosonic dual-rail representation, the two basis states considered here
belong to the same physical qubit, atom, or site.
}
This  scheme has been
exemplified by two-component bosons or fermions \cite{Ripoll2003, Tosta2019}, such as
in the dual-rail encoding optical photon proposal \cite{Knill2001, Chou2024},
and formulated in general \cite{WuLidar02}.  It can be also
a cold atom system trapped in optical lattices, where information can be encoded
in the intrinsic degrees of freedom of the cold atoms.

Single-qubit gate operations are generated by \cite{WuLidar02},
\begin{equation}
 {J}_{x} =  {c}_{0}^{\dagger} {c}_{1} +  {c}_{1}^{\dagger} {c}_{0}, \quad
 {J}_{z} =  {c}_{0}^{\dagger} {c}_{0} -  {c}_{1}^{\dagger} {c}_{1},
\end{equation}
and \( {J}_{y} = i/2\,[ {J}_{x},  {J}_{z}]\).
Together, these operators form an \(\mathfrak{su}(2)\) algebra,
as shown in Appendix~\ref{app:comm}, whose two-dimensional
representation is given by the Pauli matrices.
Physical realizations include cold atom systems \cite{Ripoll2003},
circuit quantum electrodynamics systems
\cite{Tsunoda2023} and superconducting cavities etc.
\cite{Kubica2023,Lukens2017,Ilves2020,Scala2024}. 

With the encoding and controllable generators in place, we now summarize the error-correction properties of the IPQ and provide comparisons with standard bosonic encodings.

\subsection{Error-correction properties of the IPQ}
Quantum error correction codes (QECCs) are designed based on the first-order qubit-bath interaction~\cite{Shor1995}. The total qubit-bath Hamiltonian is given by
\( {H} =  {H}_{S} +  {H}_{B} +  {H}_{SB},\)
where \( {H}_{S}\) and \( {H}_{B}\) represent the qubit and bath
Hamiltonians.  
In general, the interaction
\(H_{SB} = \vec{\sigma} \cdot \vec{B}\) is the lowest/first order of the
qubit-bath interaction, which results in Pauli errors. 
The interaction is assumed to be dominant in the design of quantum error correction schemes
\cite{Anderson2021, Neeve2022,Google2024}. The second order interactions are described by a second-rank tensor \cite{WuL02}.

Analogous to the expansion in terms of Pauli operators, the lowest order expansion of \(H_{SB}\) for an IPQ is given by: 
\begin{equation}
 {H}_{SB} = \sum_{i=0,1}\left( {c}_{i}^\dagger {B}
+ {c}_{i}  {B}^\dagger \right). \label{eq:HSB}
\end{equation}
\textcolor{black}{Here \(c_0\) and \(c_1\) denote the two internal modes of the same IPQ.
For the first-order IPQ--bath interaction to describe relaxation and mode
mixing between these two internal states, the environment must act coherently
on both modes through the same bath. This situation is physically distinct
from the individual decoherence of two separate qubits, atoms, or sites,
described by independent interactions such as
\(\vec{\sigma}_1\cdot\vec{B}_1\) and
\(\vec{\sigma}_2\cdot\vec{B}_2\).}

The second order in bosonic expansion may correspond to the conventional qubit-bath interaction
\(\vec{J} \cdot \vec{B'}\) or in the conventional notation $ \vec{\sigma} \cdot \vec{B}$, which essentially results from the dipole approximation and $B'$ has the same form as $B$.
In the present theoretical framework, the Pauli errors involve three bosonic or fermionic operators, for example,
\({c}_{0}^{\dagger}  {c}_{1} ( {b}_{\alpha}^{\dagger} +  {b}_{\alpha}),\)
which is a {\it nonlinear} term and has the same form and order as spontaneous parametric down-conversion in quantum optics.
Therefore the first  order expansion Eq. \eqref{eq:HSB} is supposed to be significantly stronger than the second  order expansion. \textcolor{black}{ The primary and secondary error structures of the conventional qubit--bath and IPQ--bath interactions are summarized in Table~\ref{tab:ECC}. }

\begin{table}
\caption{Error Classification and Comparison\label{tab:ECC}}
\begin{centering}
\begin{tabular}{ c c c c }
\hline
\hline
Model & Primary errors &  Secondary errors & Pauli errors \tabularnewline
\hline
Qubit-Boson & $\vec\sigma\cdot \vec B$ & $\vec\sigma_i \cdot \mathbf{G}_{ij} \cdot \vec\sigma_j$ & $\vec \sigma$ \tabularnewline
IPQ-Boson & $(  c_0^\dagger +  c_1^\dagger) b+h.c.$ & $\vec J\cdot \vec B$ & $\vec J$ \tabularnewline
\hline
\hline
\end{tabular}
\par\end{centering}
\end{table}

Analogous to the standard QEC theory, which is based on the first-order interaction $ \vec{\sigma} \cdot \vec{B}$,  {the codewords for IPQs should be specifically designed to address the first-order IPQ-B interaction~(\ref{eq:HSB}). Theoretically it is more similar to the simplest binomial code~\cite{PRX} than the conventional multi-qubit QECCs}. A standard QECC facilitates the detection and correction of quantum errors by encoding quantum information into a protected subspace \cite{Terhal2015}. Its main purpose is to recover the quantum state to its original form after errors occur.

Consider the dipole-dipole interaction, the system-bath interaction
simplifies  to \( {H}_{SB} = ( {x}_{1} +  {x}_{0})  {B}\) with \( {x}_i = ( {c}_i^{\dagger} +  {c}_i)/{\sqrt{2}}\).
{In this case, we will demonstrate that the simplest QECC corresponds directly to the IPQ itself, provided that conventional correction and restoration are generalized beyond unitary operations to employ physically implementable reversal operations.  }

First, the error syndromes are detected by the
parity or Stabilizer generator \(\textcolor{black}{(-1)^{n_0+n_1+1}}\), where
\( {n}_{i} =  {c}_{i}^{\dagger}  {c}_{i}\), representing the parity
measurement performed on the IPQs \cite{Lin2020}. {Unfortunately, unlike the simplest binomial encoding, the codewords $|0\rangle$ and $|1\rangle$ do not satisfy the QEC criteria ( Knill-Laflamme conditions) ~\cite{PRX} and correction cannot be made by unitary recovery. We now propose a different method for correction and restoration. It is straightforward that no correction to the IPQ system is required when the
parity measurement indicates odd parity. }
Otherwise when the parity is even, the state with error will be  \(\varepsilon\, x|\psi\rangle\)  and { cannot be corrected using the standard QEC method, where
\(\textcolor{black}{|\psi\rangle=\alpha|0\rangle+\beta|1\rangle}\)
is an arbitrary IPQ  state and \(x = x_1 + x_0\)
is the error.
\textcolor{black}{
As any QECC is designed for a specified error model, the present IPQ-based
QECC is constructed specifically for the error set
\(\{I,x_0+x_1\}\). Here, \(\varepsilon\) is an effective error strength obtained
through standard calibration or randomized benchmarking~\cite{Magesan2011}.
 As in conventional QECCs, this information is
needed to determine whether error correction is beneficial. A small
calibration error leads only to a gradual degradation of the recovery
performance.}

To address this, we now propose an innovative scheme, using a well-developed 
technique~\cite{WuLi1, WuLi2},} for error correction by introducing an ancillary 
qubit initialized in the state \(|0_q\rangle\)
such that the total state is given by their tensor product  \(\varepsilon\, x|\psi\rangle|0_q\rangle\).
Next, we apply the following unitary transformation on the combined system
\(
U = \frac{1}{\varepsilon\, x}\left(|0_q\rangle\langle 0_q| +|1_q\rangle\langle 1_q|\right)
+ \sqrt{1 - \frac{1}{\varepsilon^2\, x^2}}\left(|1_q\rangle\langle 0_q| - |0_q\rangle\langle 1_q|\right),
\)
and then measure the ancillary qubit.
If the measurement of the ancillary qubit (MAQ) yields \(|0_q\rangle\), then
\(
\langle 0|U|0\rangle \varepsilon x |\psi \rangle = |\psi\rangle,
\)
indicating that the error is corrected. If the MAQ is \(|1_q\rangle\), the IPQ state
 becomes \(\langle 1|U|0\rangle \, \varepsilon \, x \, |\psi\rangle =
i \sqrt{1 - \varepsilon^2 x^2} |\psi\rangle.\)
For small \(\epsilon\), we have
\begin{equation}
\langle1_q|U|0_q\rangle\,\varepsilon x|\psi\rangle
=
i|\psi\rangle+O(\varepsilon^2),
\end{equation}
which means that the first-order error is also corrected in this branch.
This makes the scheme a
win-win measurement (WWM), as either measurement outcome can effectively
correct the error.  Essentially,  the proposed QEC is equivalent to applying the
 inverse transform $\frac{1}{x}$ on the error states, enabling immediate recovery
  since $\frac{1}{x} x |\psi\rangle=|\psi\rangle$, where the inverse  recovery
  $\frac{1}{x}$ is implemented by the joint unitary operation $U$ plus ancillary
  measurement.

\textcolor{black}{ The above construction defines the ideal target recovery map.
In practice, the parity measurement, ancillary measurement, and recovery operation
\(U\) may all be imperfect, and their errors must be included in the platform-specific
break-even comparison between the protected and bare IPQs~\cite{Egan2021}. }



As for the break-even point, our scheme is similar to the bosonic binomial codes, where {\em beating the break-even point} is defined as achieving a logical qubit lifetime longer than the best available physical qubit lifetime and demonstrated experimentally in ~\cite{Binormial2023} and \cite{Brock2025}.  In our case, where the IPQ serves as both the physical and logical qubit, beating the break-even point means surpassing the intrinsic lifetime of the IPQ itself.  In addition, errors introduced by the measurement of the ancillary qubit must also be taken into account.
\textcolor{black}{The resulting error-detection and recovery procedure is summarized in
Table~\ref{tab:The-QEC-code}.}



\begin{table}
\caption{The QEC code with single qubit.\label{tab:The-QEC-code}}
\begin{centering}
\begin{tabular}{ c c c c  }
\hline
\hline
\text{Measure Result}  & \text{Error} & \text{Recovery}& \text{Method}
\tabularnewline
\hline
 +1 & \text{none} & $I$&\\
 -1&${x}_{1}+ {x}_{0}$ & MAQ&\text{WWM}
\tabularnewline
\hline
\hline
\end{tabular}
\par\end{centering}
\end{table}

\SMstart

\subsection{Comparison with binomial bosonic codes}
\label{sec:binomial_comparison}

We briefly compare the IPQ code with a representative binomial bosonic code
in terms of their physical Hilbert spaces, dominant errors, syndrome
measurements, and recovery procedures. The detailed binomial-code construction
is given in Appendix~\ref{app:binomial}.

A binomial code encodes a logical qubit in selected higher Fock states of a
single bosonic mode. For the representative single-photon-loss code, the
system--bath interaction is
\[
H_{\rm SB}^{B}
=
\hat a B^\dagger+\hat a^\dagger B,
\]
where \(\hat a\) is the annihilation operator of the single bosonic mode.
A photon-loss event changes the photon-number parity and maps the logical
codewords to an orthogonal error subspace. The code satisfies the
Knill--Laflamme condition for this error set and therefore permits a
conditional unitary recovery.

In contrast, the IPQ is defined in a fixed-one-particle,
two-internal-state subspace. Its dominant first-order IPQ--bath error is
\[
x=x_0+x_1.
\]
This error changes the total-number parity but does not satisfy the
Knill--Laflamme condition for the IPQ codewords. It is therefore detected by
the parity measurement and corrected through the measurement-assisted inverse
operation introduced in Sec.~II\,B, rather than by a conventional unitary
recovery. Pauli-type logical errors arise from the higher-order IPQ--bath
interaction and are therefore weaker than the leading first-order error.

Thus, although both constructions use bosonic degrees of freedom and parity
measurements, they employ different physical encodings and recovery
mechanisms. The binomial code uses higher Fock states of a single mode,
whereas the IPQ uses the two internal states of a physically implementable
qubit. The latter retains the central advantage emphasized in this work:
the same physical IPQ also serves as the QECC for its specified
collective/coherent error.

The present work differs from the three-qubit
collective-decoherence code considered in
Ref.~\cite{Yang2001}, where three distinct qubits are
involved. Related noise challenges arising from the
coupling of quantum systems to bosonic environments
have been studied in spin-boson and multilevel
superconducting-qubit settings~\cite{Otterpohl2022,Kaur2021}.
However, these works do not specifically address
collective/coherent decoherence between the two internal
modes of a single IPQ or identify the physical IPQ itself
with a QECC for this error.

\SMend

\subsection{Toward the break-even point with identical-particle qubits}
The notion of a \textit{break-even point} in quantum error correction refers to the condition where the lifetime of a protected logical qubit equals or surpasses that of the best physical qubit in the same platform,
\begin{equation}
\tau_{\mathrm{L}} \ge \tau_{\mathrm{P}},
\end{equation}
marking the onset where active error correction provides a net benefit rather than additional overhead.
In bosonic codes such as cat and binomial encodings, this threshold has been reached only marginally---for instance, the experiment of Grimm \textit{et al.} achieved $\tau_{\mathrm{L}} / \tau_{\mathrm{P}} \simeq 1.16$---because their dominant noise channels arise \emph{within} the encoded subspace through spin-like couplings of the form
\begin{equation}
H_{SB}^{(\mathrm{spin})} = \sum_i \boldsymbol{\sigma}_i \cdot \mathbf{B}_i,
\end{equation}
which act directly on logical degrees of freedom.
Consequently, even perfect correction of photon-loss events cannot eliminate logical bit-flip and dephasing errors that originate intrinsically in the microscopic Hamiltonian.

In contrast, the identical-particle qubit (IPQ) framework introduced here fundamentally alters this microscopic structure.
The first-order system--bath coupling takes a \emph{collective bilinear} form, i.e., the first term in Eq.(\ref{eq:HSB})
which preserves the logical basis and shifts all leading logical errors to higher orders in the coupling strength.
This microscopic suppression, combined with the measurement-assisted nonunitary recovery and the leakage-elimination operator (LEO) protection, effectively removes the primary logical error channel rather than merely correcting its consequences.

Therefore, the IPQ scheme provides a \emph{hardware-level route} toward surpassing the break-even point.
Because logical coherence is preserved already at the interaction level, the logical lifetime $\tau_{\mathrm{L}}$ can, in principle, exceed the physical particle lifetime even \emph{without} repeated parity-check cycles.
The unified framework of QECC, DD, and DFS developed here allows the break-even threshold to be crossed \emph{passively}, through intrinsic error prevention instead of frequent active correction.
This conceptual shift---from ``detect and correct'' to ``prevent and preserve''---represents a distinct pathway to scalable, fault-tolerant quantum computation with identical particles.

\section{Exact System-Bath Dynamics}\label{sec:exact}

We now turn to the dynamics of the encoded modes coupled to bosonic environments. Our goal is twofold: (i) to obtain closed-form expressions for the Heisenberg evolution of the relevant system operators, which directly determine the reduced dynamics in the encoding subspace, and (ii) to use these expressions to design control protocols that suppress leakage and logical errors. In the main text we present the central results and their physical consequences; detailed derivations are deferred to Appendix~\ref{app:collective}.

\subsection{Hamiltonian of the IPQ--boson model}
The identical particles forming the IPQ interact with a common bosonic thermal reservoir
at a finite inverse temperature \(\beta\) ,or couple to their individual bosonic reservoirs at finite temperatures
\(T_{i}\). The total Hamiltonian is given by
\[
 {H}_{\text{tot}}= {H}_{\text{S}}+ {H}_{\text{B}}+ {H}_{\text{SB}},
\]
where \( {H}_{\text{B}}=\sum_{i,\alpha}\omega_{i,\alpha} {b}_{i,\alpha}^{\dagger} {b}_{i,\alpha}\)
represents the reservoir Hamiltonian for individual decoherence, and
\( {H}_{\text{B}}=\sum_{\alpha}\omega_{\alpha} {b}_{\alpha}^{\dagger} {b}_{\alpha}\)
describes the collective decoherence reservoir Hamiltonian. The interaction Hamiltonian is
\[
 {H}_{\text{SB}}=\sum_{i} {L}_{i}\otimes {B}_{i}+\text{h.c.},
\]
where \( {L}_{i}\) and \( {B}_{i}\) are the system and reservoir operators, respectively.
Here, we consider the rotating wave approximation for \( {H}_{\text{SB}}\).

For collective reservoir coupling, there is only one system operator involved in \( {H}_{\text{SB}}\),
i.e., \( {L}_{0}=\left( {c}_{0}+ {c}_{1}\right)/\sqrt{2}\), with the corresponding reservoir
operators given by \( {B}_{0}=\sum_{\alpha}g_{\alpha}^{*} {b}_{\alpha}^{\dagger}\).
In contrast, if the identical particles interact with their reservoirs individually, the interaction
Hamiltonian is
\[
 {H}_{\text{SB}}=\sum_{i=0,1} {c}_{i}\otimes\sum_{\alpha}g_{i,\alpha}^{*} {b}_{i,\alpha}^{\dagger}+\text{h.c.}
\]

The IPQ Hamiltonian consists of two parts, \( {H}_{\text{S}}= {H}_{\text{S}}^{0}+ {H}_{\text{C}}\), where
\[
 {H}_{\text{S}}^{0}=\sum_{k=x,z}G_{k} {J}_{k}
\]
is the gate operation Hamiltonian, and
\begin{equation}
 {H}_{\text{C}}=\mu(t)\sum_{i} {c}_{i}^{\dagger} {c}_{i} \label{eq:hc}
\end{equation}
is the  LEO Hamiltonian.

\subsection{Collective decoherence}
In this section, the identical particles couple to a common thermal reservoir, which
induces collective decoherence in the IPQ system. To simplify the analysis, we introduce
 two collective operators: \( {a}_{1} = \left( {c}_{1} +  {c}_{0}\right)/\sqrt{2}\) and
\( {a}_{0} = \left( {c}_{1} -  {c}_{0}\right)/\sqrt{2}\), which satisfy the commutation
relation \(\left[ {a}_{i},  {a}_{j}^\dagger\right] = \delta_{ij}\).
\textcolor{black}{ Here \(a_0\) and \(a_1\) are collective linear combinations
of the two IPQ modes \(c_0\) and \(c_1\). They are unrelated to the single-mode
annihilation operator \(\hat a\) used in the binomial-code example in
Appendix~\ref{app:binomial}. }
In terms of these collective operators, the interaction Hamiltonian becomes
\[
 {H}_{\text{SB}} = \sum_{\alpha}g_{\alpha}^{*}  {a}_{1} \otimes  {b}_{\alpha}^{\dagger} + \text{h.c.},
\]
where \( {a}_{1}\) interacts with the reservoir modes.
The corresponding gate operation operators can be expressed in terms of \(a_i\) as follows
\[
 {J}_{x} =  {a}_{1}^{\dagger}  {a}_{1} -  {a}_{0}^{\dagger}  {a}_{0},
\]
\[
 {J}_{z} =  {a}_{1}^{\dagger}  {a}_{0} +  {a}_{0}^{\dagger}  {a}_{1}.
\]
Additionally, the LEO Hamiltonian [Eq. (\ref{eq:hc})] in this representation reads
\[
 {H}_{\text{C}} = \mu(t) \sum_{i=0,1}  {a}_{i}^{\dagger}  {a}_{i}.
\]
This structure allows for the application of LEO pulses to protect the system from the effects of decoherence.

To obtain the exact dynamics under collective decoherence, we solve the Heisenberg equations of motion for the system operators and eliminate the reservoir modes. The full derivation and the explicit time-dependent coefficients are given in Appendix~\ref{app:collective}. Here we only quote the resulting operator solution,
\begin{equation}
a_i(t)=\sum_{k=0,1} C_i^{k}(t)\,a_k(0)+\sum_{\alpha} B_{c,\alpha}^{i}(t)\,b_{\alpha}(0),
\label{eq:collective_solution}
\end{equation}
which fully determines all system observables for arbitrary initial system and bath states.

The IPQ-B interaction can be written as
\( {H}_{\text{SB}}=\sum_{\alpha}g_{\alpha}\left( {a}_{1}^{\dagger}
 {b}_{\alpha} +  {a}_{1}  {b}_{\alpha}^{\dagger}\right)\), where
\( {a}_{1}=( {c}_{1}+ {c}_{0})/\sqrt{2}\) and the commuting system operator
is \( {a}_{0}=( {c}_{1}- {c}_{0})/\sqrt{2}\).
The system Hamiltonian is given by
\( {H}_{\text{S}}^{0}=\sum_{k=0,x,z}G_{k} {J}_{k}\), with \(G_{k}\)
 representing the gate operation strength. Here, \( {J}_{0}= {I}_{2}\)
 denotes an identity operator corresponding to the information storage process.
The Hamiltonian for the thermal bosonic reservoir is
 \( {H}_{\text{B}}=\sum_{\alpha}\omega_{\alpha} {b}_{\alpha}^{\dagger} {b}_{\alpha}\).

We expand the identical-particle operators at time \(t\) in the initial system/bath modes,
\begin{equation}
  {c}_{j}(t)=\sum_{i=0,1} C_{i}^{j}(t)\,{c}_{i}^{0}
           +\sum_{\alpha} B_{c,\alpha}^{j}(t)\,{b}_{\alpha}^{0},
  \label{eq:exps}
\end{equation}
In particular, under collective decoherence the shared reservoir induces mode mixing between the two identical-particle modes. For example, setting \(j=0\) in Eq.~(\ref{eq:exps}) gives
\begin{equation}
  {c}_{0}(t)= C_{0}^{0}(t)\,{c}_{0}^{0}+ C_{1}^{0}(t)\,{c}_{1}^{0}
           +\sum_{\alpha} B_{c,\alpha}^{0}(t)\,{b}_{\alpha}^{0},
  \label{eq:cc0}
\end{equation}
so that \( {c}_{0}(t)\) generally depends on both \( {c}_{0}^{0}\) and \( {c}_{1}^{0}\).

\begin{figure}
\includegraphics[scale=0.50]{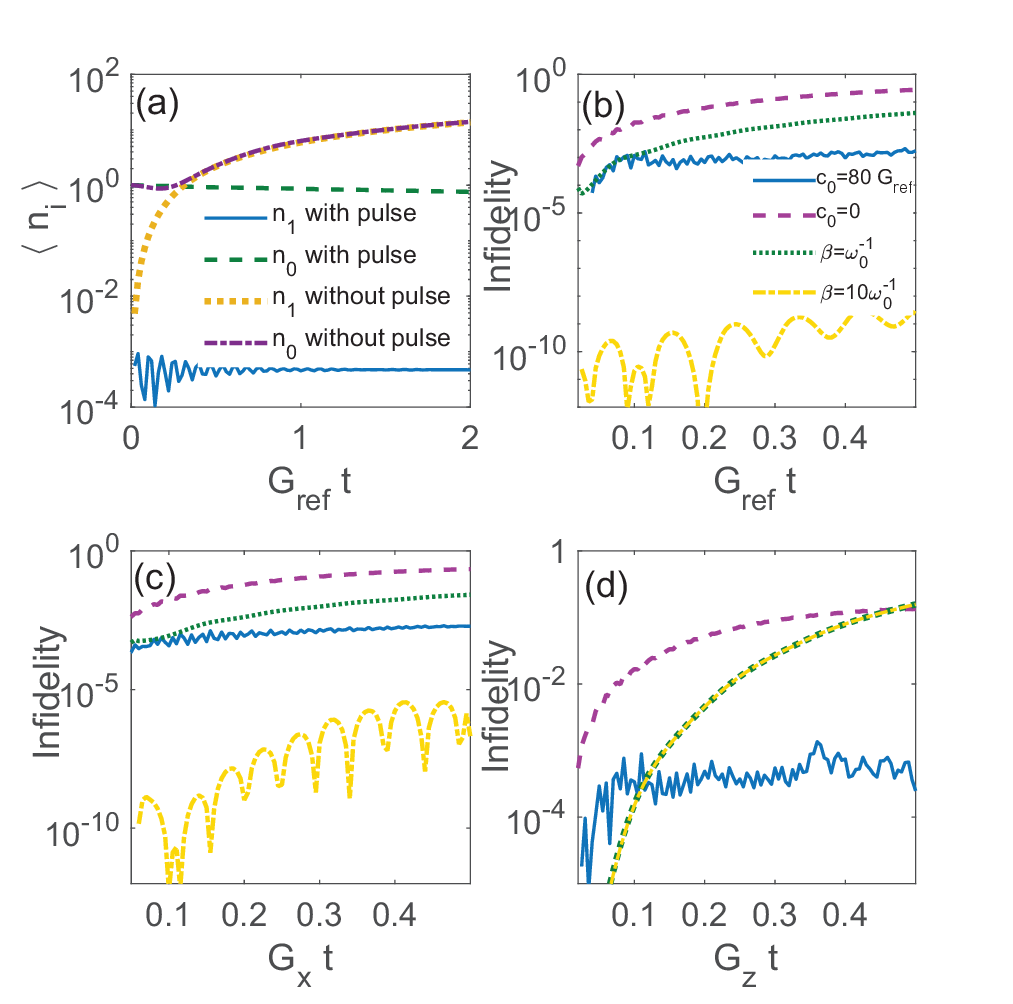}
\caption{(a) Time evolution of the mean excitation numbers
\(\langle n_0(t)\rangle\) and \(\langle n_1(t)\rangle\) for a single IPQ,
with and without LEO pulses.
(b)--(d) Infidelity of the effective IPQ state during
(b) information storage, (c) the \(X\)-gate operation, and
(d) the \(Z\)-gate operation.
The black solid and purple dashed curves correspond to the dynamics with and
without LEO pulses, respectively. The black dotted and yellow dash-dotted
curves show the evolution from the decoherence-free state associated with
\(a_0(t)\) at
\(\beta=\omega_0^{-1}\) and
\(\beta=10\omega_0^{-1}\), respectively.
The reservoir is described by a Lorentzian spectral density with
\(\Omega=\omega_0\),
\(\Gamma=5G_{\rm ref}\), \(\gamma=0.5G_{\rm ref}\), and
\(\omega_0=100G_{\rm ref}\).
The LEO pulse amplitude, width, and spacing are \(c_0=80G_{\rm ref}\), \(\Delta\tau=0.02\pi G_{\rm ref}^{-1}\), and \(\delta\tau=0.005\pi G_{\rm ref}^{-1}\), respectively. Here, \(G_{\rm ref}\) is an arbitrary reference frequency used to nondimensionalize time and all frequency parameters; it is not a gate coupling. We set \(G_{\rm ref}=1\) in the numerical calculations. For the \(X\)- and \(Z\)-gate operations, the gate strengths are chosen as \(G_x=G_z=G_{\rm ref}\), whereas no gate operation is applied during information storage.}
\label{fig:n_inf}
\end{figure}

Without additional assumptions (the Appendix treats more general bilinear cases beyond the rotating-wave approximation), we derive exact evolution equations for the coefficients.
These equations directly yield single- and two-particle observables, as detailed in Appendices~\ref{app:collective} and~\ref{app:initial}, and the model is analytically solvable~\cite{Wang2024}.
For individual dynamics we evaluate the average excitation
\( {n}_{i}(t)={c}_{i}^{\dagger}(t){c}_{i}(t) \),
\[
\langle {n}_{i}(t)\rangle
=\sum_{j=0,1}\!\!\left|C_{j}^{i}(t)\right|^{2}
   \langle {c}_{j}^{0\dagger}{c}_{j}^{0}\rangle_{\mathrm{S}}
 +\sum_{\alpha}\!\left|B_{c,\alpha}^{i}(t)\right|^{2}
   \frac{1}{e^{\beta\omega_{\alpha}}-1}.
\]
As \(t\!\to\!\infty\), \(|C_{i}^{i}(t)|^{2}\!\to\!0\) and the single particle thermalizes to the bath temperature, consistent with the yellow dotted and purple dot-dashed curves in Fig.~\ref{fig:n_inf}(a).

To assess gate errors we define the target-IPQ effective state
\[
 {\rho}_{\mathrm{eff}}
 = \tfrac{1}{2}\!\left( {I}_{2}
 +\!\sum_{k=x,y,z}\!\langle {J}_{k}\rangle\,{\sigma}_{k}\right),
\]
obtained by tracing out all non-target qubits from the (possibly entangled) multi-IPQ state at the evaluation time. The effective Bloch components are
\[
\langle {J}_{k}(t)\rangle
= \sum_{i,j=0,1}\!\mathcal{C}_{ij}^{k}(t)
  \langle {c}_{i}^{0\dagger}{c}_{j}^{0}\rangle_{\mathrm{S}}
 +\sum_{\alpha}\!\mathcal{B}_{\alpha}^{k}(t)\,
  \frac{1}{e^{\beta\omega_{\alpha}}-1},
\]
where \(\mathcal{C}_{ij}^{k}(t)\) and \(\mathcal{B}_{\alpha}^{k}(t)\) are functions of
\(C_{i}^{k}(t)\) and \(B_{c,\alpha}^{k}(t)\), as derived in Appendix~\ref{app:collective}.
The first term captures gate-operation, initial-state, and dissipative contributions \cite{Burgarth2021};
the second is the purely thermal contribution (the only part depending on \(1/\beta\)).

\subsection{Dynamical decoupling with LEO}
We first determine the LEO operator and explore how it dynamically decouples 
the interaction between the IPQ and bath~\cite{Wu2002,Suter2016}.
By following the procedure in Ref.  \cite{Wu2002}, it is easy to show:  
\( {R}_{L} =  {R}_{L}^{\dagger} =
 e^{-i\pi( {c}_{0}^{\dagger} {c}_{0} +  {c}_{1}^{\dagger} {c}_{1})}\),
 such that \( {R}_{L} {H}_{SB} {R}_{L} = - {H}_{SB}\) and
 \( {R}_{L} {H}_{a} {R}_{L} =  {H}_{a}\), where \(a = S, B\).
 Ideally, by toggling \( {R}_{L}\) on and off over a duration of \(2\delta t\),
 the entire system evolves as follows:
\(
e^{-i {H}\delta t} {R}_{L}e^{-i {H}\delta t} {R}_{L} = e^{-2i( {H}_{S} +  {H}_{B})\delta t}
\)
, indicating that ${R}_{L}$ acts as an LEO with the corresponding Hamiltonian
\( {H}_{\text{LEO}}(t) = \mu(t)\sum_{i=0}^{1} {c}_{i}^{\dagger} {c}_{i}\)
\cite{Wu2002,Zheng2020}.
\(\mu(t)\) is an external field on the system, for example, an external
magnetic field such that \(\mu_{\text{eff}} = \mu + g\mu_{b}|B_{z}|/2\)   \cite{NP}.
By applying a time-dependent \(B_{z}\), one can achieve a controllable
\(\mu_{\text{eff}}\). It is noteworthy that the LEO is effective for all linear couplings
\( {H}_{\text{SB}}\). Significantly,
the LEO Hamiltonian \( {H}_{\text{LEO}}\) commutes with the set
\(\{ {J}_{k}\}\) and all single- and double-qubit gates, allowing
\( {R}_{L}\) to be applied independently of these gates as background
pulses or noise.

To clarify the mechanism by which LEO pulses
suppress decoherence, we focus on the subspace of operators constructed from the
initial system annihilation operators \(\{ {c}_{0}^{0},  {c}_{1}^{0}\}\), referred
to as \emph{the encoding subspace}. As observed from Eq. (\ref{eq:exps}), if the system
is closed, not only will the system annihilation operators at any given time remain
within the encoding subspace,
but the sum of the squared magnitudes of the expansion coefficients in Eq. (\ref{eq:exps})
will also remain constant. Within this encoding subspace, there are four independent
expansion coefficients. We define a vector of coefficients as
\(\vec{C}(t) = [C_{1}^{1}(t), C_{0}^{1}(t), C_{1}^{0}(t), C_{0}^{0}(t)]^{\text{T}},\)
which encompasses all the relevant components.
 As derived in Appendix~\ref{app:collective}, the dynamical
 equation for \(\vec{C}(t)\), encompassing all operations executed
 on the IPQ, can be expressed in the unified form
 \begin{eqnarray}
 \dot{\vec{D}}_k(t) &=&-i\,G_k\bar{\sigma}_{z}\otimes\bar{I}_{2}\vec{D}_k(t)  \label{eq:Dt} \\
 &&-\int_{0}^{t}\mathrm{d}s\, \exp\left[-i\int_{s}^{t}\mathrm{d}\tau\,\mu(\tau)\right]
 \bar{F}(t-s)\vec{D}_k(s), \nonumber
 \end{eqnarray}
where \(\vec{D}_k(t) = \bar{U}_{k} \vec{C}(t)e^{-i\,\int_{0}^{t} \text{d}\tau\,\mu(\tau)}\)
is a slow-varying vector with a diagonalized gate operation matrix
\(\bar{\sigma}_{z} \otimes \bar{I}_{2}\). Here, \(k = x, z, 0\) corresponds to the
X gate operation, the Z gate operation, and the information storage process
(with \(G_{0} = 0\)), respectively. The matrix \(\bar{U}_{k}\) is a constant matrix.
The function \(\bar{F}(t-s)\) is an integral-kernel matrix containing the reservoir
correlation functions, whose explicit form is given in Appendix~\ref{app:collective}.
When \( e^{-i\int_{t}^{t'}\mu(\tau)d\tau} \) represents a fast oscillation term and
\( \bar{F}(t-s) \) is a slowly varying matrix, the integral term in Eq. (\ref{eq:Dt})
approaches zero according to the Riemann-Lebesgue lemma \cite{Gradshteyn1980,Brown1993}.
As a result, Eq. (\ref{eq:Dt}) simplifies to
\(\dot{\vec{D}}(t) = -i\,G_{k}\bar{\sigma}_{z} \otimes \bar{I}_{2} \vec{D}(t),\)
which corresponds to the ideal \( k \)-gate operation for the IPQ.

Our numerical analysis is shown in FIGs. \ref{fig:n_inf} (a-d), where the system initial state is
 \( |\psi_{0}\rangle = \frac{\sqrt{2}}{2}(|1\rangle + |0\rangle) \). We
employ the Lorentzian spectral density given by
\(J(\omega) = \frac{\Gamma \gamma^2/2\pi}{(\omega - \Omega)^2 + \gamma^2}\)
where \( \gamma \) represents the spectral width, \( \Omega \) denotes the central
frequency of the reservoir, and \( \Gamma \) is the global dissipation rate. Numerical results further confirm that the application of LEO pulses effectively suppresses single-particle thermalization, allowing the particles to maintain their original quantum states, as shown by the solid black line and dashed black line in
FIG. \ref{fig:n_inf} (a). Then we examine the dynamics of infidelity (error rate)
during gate operations. As illustrated by the purple dashed line in
FIG. \ref{fig:n_inf} (b-d), the infidelity between \( \rho_{\text{eff}} \)
and \( \rho_{\text{eff}}^{0} \) rapidly increases when LEO pulses are
absent, where \( \rho_{\text{eff}}^{0} \) is the effective density
matrix without the IPQ-B interaction. In contrast, when LEO pulses
are implemented, the infidelity across various gate operations remains
below \( 10^{-3} \), as shown by the solid black lines in FIG. \ref{fig:n_inf} (b-d).

The collective mode \(a_{0}(t)\) is invariant because \([a_{0},H_{SB}]=0\).
The eigenstate of \(a_{0}^{\dagger}(0)a_{0}(0)\),
\(|\psi_{\text{DFS}}\rangle=\tfrac{\sqrt{2}}{2}(|1\rangle-|0\rangle)\),
thus forms a decoherence-free subspace (DFS) usable as a special initial state.
Without LEO, Fig.~\ref{fig:n_inf}(b-d) shows black dotted (\(\beta=\omega_{0}^{-1}\))
and yellow dot-dashed (\(\beta=10\omega_{0}^{-1}\)) traces: storage and \(X\) gates
improve at lower temperature, whereas the \(Z\)-gate infidelity is essentially
temperature-insensitive. With LEO, the \(Z\) gate remains high-fidelity and thermal
noise is strongly suppressed for all gates. Further analytical derivations are
provided in Appendix~\ref{app:collective}.

\begin{figure}
\includegraphics[scale=0.3]{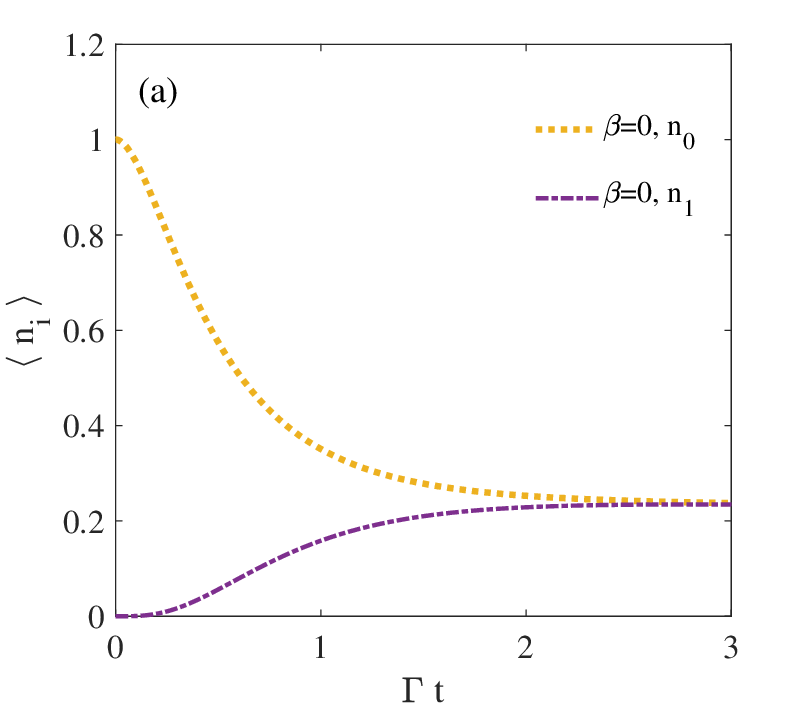}
\includegraphics[scale=0.3]{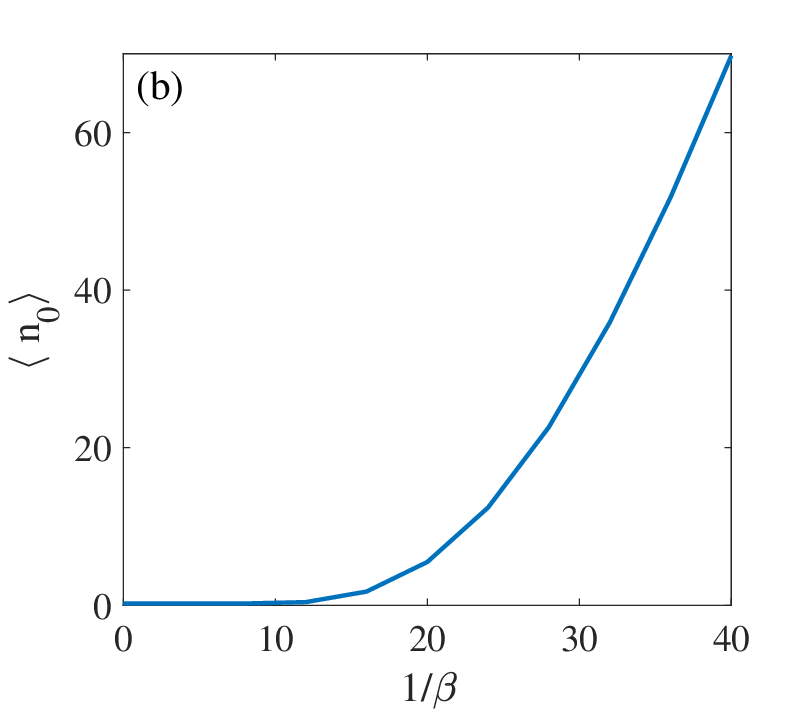}
\caption{
Common-bath occupation dynamics in the absence of gate operations.
(a) Average excitation numbers as functions of the dimensionless time
\(\Gamma t\).
(b) Steady-state average excitation number as a function of \(1/\beta\).
The system parameters are \(\gamma=2.5\Gamma\) and
\(\omega_0=100\Gamma\). All other parameters are in units of \(\Gamma=1\).
}
\label{fig:AENT}
\end{figure}

\SMstart
We also examine the occupation dynamics in the common-bath case without gate
operations. The IPQ is initially prepared with one particle in internal mode
\(0\),
\[
|0\rangle=c_0^\dagger|\mathrm{vac}\rangle .
\]
As shown in Fig.~\ref{fig:AENT}(a), at zero reservoir temperature the
occupations of the two internal modes evolve toward a common steady value.
The steady state does not correspond to complete relaxation of the initial
excitation because the common reservoir mediates an effective indirect
coupling between the two internal modes. As the reservoir temperature
increases, the average occupation of each mode also increases, as shown in
Fig.~\ref{fig:AENT}(b). These results illustrate the common-bath-induced mode
mixing generated by the leading IPQ--bath interaction.
\SMend

\subsection{Two-qubit gate}
We consider a two-logical-qubit system composed of two IPQs, \(a^\alpha\) and \(a^\beta\),
where information is encoded in logical qubits as
\( |i\rangle_{k} =  {a}_{i}^{k\dagger} |\text{vacuum}\rangle \),
with qubit labels \(k = \alpha, \beta\) and state labels \( i = 0, 1 \).
A control-phase (C-phase) gate \cite{Heuck2020} can be implemented using
the Hamiltonian
\(H_{\text{cp}} = \varepsilon_{\text{cp}}  {n}_{0}^{\alpha}  {n}_{0}^{\beta},\)
where \(  {n}_{0}^{\alpha,\beta} \) are particle number operators. The corresponding
evolution operator for the C-phase gate is
\(U_{\text{cp}} = \exp \left( -i \int_{0}^{\tau} \varepsilon_{\text{cp}}(\tau)  {n}_{0}^{\alpha}  {n}_{0}^{\beta} \, d\tau \right).\)
Since the coupling only involves the quantum state \( |0\rangle_{k} \),
we have \( H_{\text{cp}} |1\rangle_{k} = 0 \) for all \( k \). This implies
that the C-phase gate does not induce any phase shift on quantum states
containing \( |1\rangle_{k} \). When \( H_{\text{cp}} \) acts on
\( |0\rangle_{\alpha} |0\rangle_{\beta} \), \( U_{\text{cp}} \) generates a phase shift of
\(\alpha_{\text{cp}} = \int_{0}^{\tau} \varepsilon_{\text{cp}}(\tau) \, d\tau\), resulting in a C-phase gate when $\alpha_{\text{cp}} =\pi$.
We designed the C-phase gate in this way because \( H_{\text{cp}} \) commutes
perfectly with the LEO Hamiltonian \( H_{\text{LEO}} \). This ensures that
applying LEO pulses to protect the C-phase gate operation does not introduce
additional errors, highlighting a key advantage of our proposal in the context
of dynamical decoupling.
Since
the C-phase gate involves only two-body interactions between the identical particles
\(  {a}_{0}^{\alpha} \) and \(  {a}_{0}^{\beta} \), the coupling strength
\( \varepsilon_{\text{cp}} \) should be comparable to that of single-qubit gate
operations. As shown in our numerical results for the single gate operations,
even when the gate operation
strength is much lower than the particle-system coupling strength, high fidelity
can still be achieved with LEO pulse application. Thus, our proposal
demonstrates excellent robustness.

Physically, interactions between different optical wells in cold atom optical lattices can be achieved by tuning atomic contact interactions or long-range dipole-dipole interactions. These are typically controlled by adjusting the atomic scattering length or using Feshbach resonances \cite{Bloch2008}.
In circuit quantum electrodynamics (cQED) systems, two superconducting resonators or qubits can interact through shared nonlinear elements, like Josephson junctions, leading to an interaction Hamiltonian of the form
\( H_{\text{cp}} \) \cite{Blais2021}. Kerr nonlinearities or cross-mode coupling are common ways to realize these interactions \cite{Aspelmeyer2014}. Additionally, in cQED, dispersive coupling under far-detuned conditions enables interactions between qubits and cavity modes that generate frequency shifts based on particle number, resulting in an effective
 \( H_{\text{cp}} \) interaction \cite{Koch2007}.

\textcolor{black}{Together with the single-IPQ generators \(J_x\) and \(J_z\), the
C-phase interaction forms a universal gate set for the IPQ. Since the physical
IPQ itself also serves as the QECC code space for the specified
coherent/collective error, the same single- and two-IPQ operations act at both
the physical-qubit and logical-QECC levels. Thus, the physical and logical
gate operations are implemented in the same manner in the present framework.}

\section{Individual decoherence}\label{sec:indiv}

\SMstart
Besides the common-bath-induced coherent error discussed above, independent
reservoirs can also induce individual, incoherent decoherence of the two
internal modes. In this section, we show that such individual-reservoir noise
can be suppressed by LEO-based dynamical decoupling. The technical derivations
are collected in Appendix~\ref{app:individual}.
\SMend

We consider the case in which the two internal modes of the IPQ couple
separately to independent bosonic thermal reservoirs. The reservoirs may have
similar spectral densities but different inverse temperatures \(\beta_i\). In
contrast to the common-bath analysis above, here we keep the counter-rotating
terms and do not make the rotating-wave approximation. The total Hamiltonian is
$$ {H}_{\text{tot}}= {H}_{\text{S}}+ {H}_{\text{B}}+ {H}_{\text{SB}},$$
where $  {H}_{\text{B}}=\sum_{i,\alpha}\omega_{i,\alpha} {b}_{i,\alpha}^{\dagger} {b}_{i,\alpha}$
is the reservoir Hamiltonian and
$$ {H}_{\text{SB}}=\sum_{i=0,1}\left( {c}_{i}^{\dagger}+ {c}_{i}\right)
\sum_{\alpha}g_{i,\alpha}\left( {b}_{i,\alpha}^{\dagger}+ {b}_{i,\alpha}\right)$$
with the coupling strengths $g_{i,\alpha}$.
The IPQ Hamiltonian contains two parts, $$ {H}_{\text{S}}= {H}_{\text{S}}^{0}+ {H}_{\text{C}},$$
where $ {H}_{\text{S}}^{0}=\sum_{i}\omega_{0,i} {c}_{i}^{\dagger} {c}_{i}+\sum_{K=x,z}G_{K} {J}_{K}$
is the free Hamiltonian and the gate Hamiltonian, and $ {H}_{\text{C}}=\mu(t)\sum_{i} {c}_{i}^{\dagger} {c}_{i}$
  is the LEO Hamiltonian.

\subsection{Model and main results}
Starting from the total Hamiltonian above and eliminating the independent reservoirs, the exact non-Markovian evolution can be written in a linear form in terms of the initial system and bath operators,
\begin{eqnarray*}
 {c}_{j}(t) & = & \sum_{i=0,1}\left(C_{i}^{j}(t) {c}_{i}^{0}+\bar{C}_{i}^{j}(t) {c}_{i}^{0\,\dagger}\right)\\
 &&+\sum_{i,\alpha}\left(B_{i,\alpha}^{j}(t) {b}_{i,\alpha}^{0}+\bar{B}_{i,\alpha}^{j}(t) {b}_{i,\alpha}^{0\,\dagger}\right),\\
 {c}_{j}^{\dagger}(t) & = & \sum_{i=0,1}\left(C_{i}^{j\,*}(t) {c}_{i}^{0\,\dagger}+\bar{C}_{i}^{j\,*}(t) {c}_{i}^{0}\right)\\
 &&+\sum_{i,\alpha}\left(B_{i,\alpha}^{j\,*}(t) {b}_{i,\alpha}^{0\,\dagger}+\bar{B}_{i,\alpha}^{j\,*}(t) {b}_{i,\alpha}^{0}\right),
\end{eqnarray*}
where we have used the definitions $ {c}_{i}^{0}\equiv {c}_{i}(0)$
and $ {b}_{i,\alpha}^{0}\equiv {b}_{i,\alpha}(0).$

For later use, it is convenient to note that, in the absence of direct mode--mode coupling (e.g., during information storage with \(G_{x}=G_{z}=0\)), the individual-reservoir structure implies that each mode couples only to its own bath. In particular,
\begin{eqnarray}
 {c}_{0}(t)&=& C_{0}^{0}(t)\,{c}_{0}^{0}+\bar{C}_{0}^{0}(t)\,{c}_{0}^{0\,\dagger}\nonumber\\
 &&+\sum_{\alpha}\!\left(B_{0,\alpha}^{0}(t)\,{b}_{0,\alpha}^{0}+\bar{B}_{0,\alpha}^{0}(t)\,{b}_{0,\alpha}^{0\,\dagger}\right),
 \label{eq:dbi0-1}
\end{eqnarray}
and analogously for \( {c}_{1}(t)\) with the replacement \(0\rightarrow 1\). This absence of reservoir-mediated cross terms is the key ingredient behind the thermal-noise cancellation discussed below.
Detailed derivations are provided in Appendix~\ref{app:individual}.

First, we consider the time evolution of the average excitation number
of a single identical particle in the absence of gate operations, as shown in Fig.~\ref{fig:AENT} (a).
We assume that initially a single excitation occupies mode~0,
i.e., $\ket{0} = c_{0}^{\dagger}\ket{\mathrm{vacuum}}$.
When the environment is at zero temperature (yellow dotted and purple dash--dotted lines),
the excitations in the two modes tend toward equilibrium.
Unlike the case of independent decoherence, the steady state under collective decoherence
does not correspond to zero excitation number.
This is because the common environment induces an effective indirect interaction between the two modes.
As the environmental temperature increases,
the average excitation number in each mode also increases [see Fig.~\ref{fig:AENT} (b)].
Therefore, the inclusion of all error channels in the leading-order interaction
is physically well justified.

\begin{figure}
\includegraphics[scale=0.3]{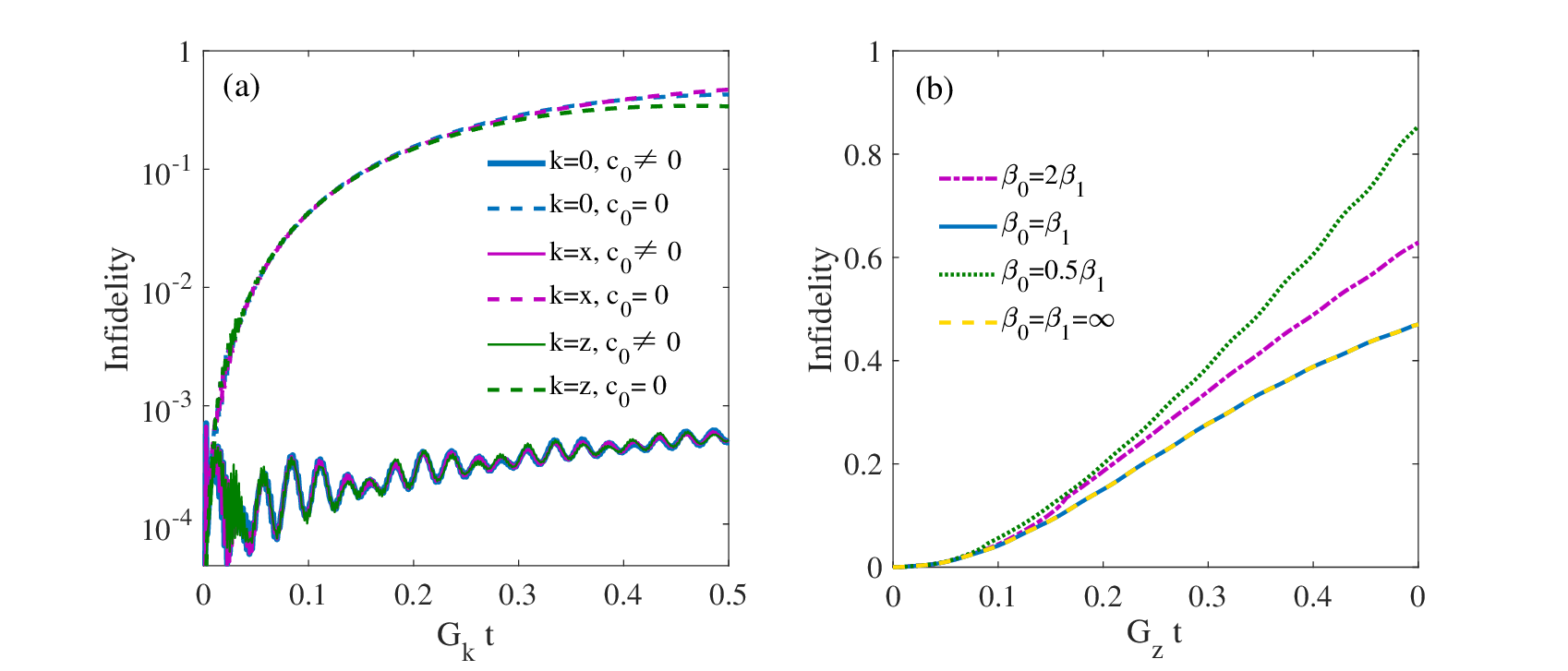}

\caption{The infidelity as a function of dimensionless time $G_{k}t$.
(a) Infidelity for different gate operations with $\beta_{1} = \beta_{2} = 0.1\omega_{0}^{-1}$.
(b) Infidelity of the \(Z\)-gate operation for different reservoir-temperature
differences in the absence of LEO pulses.
The system parameters are set to $\Gamma = 5G_{k}$, $\gamma = 0.5G_{k}$,
and $\omega_{0} = 100G_{k}$. The LEO pulse parameters include pulse strength
$c_{0} = 80G_{k}$, pulse width $\Delta\tau = \pi / 50G_{k}$, and pulse spacing
$\delta\tau = \pi / 200G_{k}$. All other parameters use $G_{k} = 1$, with $k = 0, x, z$.
\label{fig:ind_temp}}
\end{figure}

As shown in Fig. \ref{fig:ind_temp} (a), the LEO method continues to perform
exceptionally well in protecting gate operations. In contrast, when LEO pulses
are not applied, the optimal fidelity is achieved at zero temperature difference,
regardless of the environmental temperature, as confirmed by Fig. \ref{fig:ind_temp} (b).
Additionally, we observe that at any temperature, as long as the two thermal
reservoirs have no temperature difference, fidelity remains unchanged.
This indicates that under zero temperature difference conditions,
both information storage and Z-gate operations are free from thermal noise.
To make this cancellation transparent, we quote the analytic expressions for the effective Bloch components (derived in Appendix~\ref{app:individual}). In the continuous limit one finds
\begin{eqnarray}
\langle J_{z}(t) \rangle & = & \sum_{i,j=0}^{1} \left( C_{j}^{1*}(t) C_{i}^{1}(t) - C_{j}^{0*}(t) C_{i}^{0}(t) \right) \langle  {c}_{j}^{0\dagger}  {c}_{i}^{0} \rangle_{\text{S}} \nonumber\\
& & + \int d\omega\, J_{1}(\omega)\, \mathcal{B}_{1}(\omega)\, \frac{1}{e^{\beta_{1} \omega} - 1} \nonumber\\
 && - \int d\omega\, J_{0}(\omega)\, \mathcal{B}_{0}(\omega)\, \frac{1}{e^{\beta_{0} \omega} - 1},
\label{Jzind}
\end{eqnarray}
and
\begin{eqnarray}
\langle J_{x}(t) \rangle & = & \sum_{i,j=0}^{1} \left( C_{i}^{1*}(t) C_{j}^{0}(t) + C_{i}^{0*}(t) C_{j}^{1}(t) \right) \left\langle  {c}_{i}^{0\dagger}  {c}_{j}^{0} \right\rangle_{\text{S}}, \nonumber\\
\langle J_{y}(t) \rangle & = & \sum_{i,j=0}^{1} i \left( C_{i}^{1*}(t) C_{j}^{0}(t) - C_{i}^{0*}(t) C_{j}^{1}(t) \right) \left\langle  {c}_{i}^{0\dagger}  {c}_{j}^{0} \right\rangle_{\text{S}}.
\label{Jxind}
\end{eqnarray}
This observation can be explained by Eqs. (\ref{Jzind}) and (\ref{Jxind}).
Specifically, the x- and y-components of the Bloch
vector are independent of the reservoir temperatures. When the two bosonic
thermal reservoirs have the same temperature, the thermal noise
contributions to the z-component cancel each other out.

From a physical perspective, the absence of thermal noise is caused by the
interaction with individual reservoirs. When the IPQ interacts with a common
thermal reservoir, the shared environment induces indirect coupling between the
identical particles, causing \(  {c}_0(t) \) to depend not only on
the expansion coefficients of \(  {c}_0^0 \), but also on those of \(  {c}_1^0 \) [see Eq.(\ref{eq:cc0})].
In contrast, when identical particles interact with their own thermal reservoirs,
\(  {c}_0(t) \) is expanded only in terms of \(  {c}_0 \), isolating the
transitions between the identical particles [see Eq.(\ref{eq:dbi0-1})]. When the thermal noise effects are
identical, they cancel each other out, rendering the system's dynamics unaffected
by environmental temperature. As shown in Fig. \ref{fig:ind_temp} (a), under the
influence of LEO pulses, the impact of thermal noise is significantly suppressed
across all gate operations.

\section{Conclusion}

We have shown that the first-order
system-bath interaction in our identical-particle qubit scheme
differs fundamentally from conventional first-order qubit-bath
interactions, as illustrated in Table \ref{tab:ECC}.
This new observation requires traditional strategies for combating
decoherence, including QECC, DD, and DFS, to be reformulated
while retaining their fundamental principles.
 Remarkably, we demonstrate
that these strategies can be unified into a single code, specifically
an unconventional quantum error-correcting code, provided that
conventional recovery procedures are generalized beyond unitary
operations to include physically implementable reversal operations,
\textcolor{black}{thereby placing a physically implementable qubit and a quantum
error-correction code against collective decoherence between the two bosonic
modes on equal footing.}

We also find that the stabilizer generator is consistent with the LEO
operation and that one of the IPQ states forms a DFS.  Interestingly,
the LEO in our framework is also applicable to conventional qubit--bath interactions.
 Moreover, well-established quantum error correction (QEC)
 schemes must also be employed when these second-order IPQ-bath
 interactions in Table \ref{tab:ECC} become comparatively
 significant.
The analytically solvable IPQ--bath model also enables a precise assessment
of existing protocols for combating decoherence, in particular
numerical demonstrations of the effectiveness of the LEO protocol.

\begin{acknowledgments}
This work was supported by the National Natural Science
Foundation of China (NSFC) under Grants Nos. 12574384,
12375009, 12075050 and 12205037; by the Liaoning
Provincial Science and Technology Joint Program (Key
Research and Development Program Project) under Grant No.
2025JH2/101800316;  by the grant PID2021-126273NB-I00 funded by MCIN/AEI/10.13039/501100011033, and by {\em ERDF A way of making Europe} and the Basque Government through Grant No. IT1470-22. This work has been financially supported by the Ministry for Digital Transformation and of Civil Service of the Spanish Government through the QUANTUM ENIA project call- Quantum Spain project, and by the European Union through the Recovery, Transformation and Resilience Plan-NextGenerationEU within the frame- work of the Digital Spain 2026 Agenda.  L.-A.W. thanks Prof. D. Lidar for early communication on this paper.

\end{acknowledgments}

\appendix
\onecolumngrid

\section{Commutation relations of the single-qubit generators}\label{app:comm}

The single-qubit gate generators are defined as
\begin{align*}
J_x&=c_0^\dagger c_1+c_1^\dagger c_0,\\
J_y&=-i\left(c_0^\dagger c_1-c_1^\dagger c_0\right),\\
J_z&=c_0^\dagger c_0-c_1^\dagger c_1.
\end{align*}
Using the bosonic commutation relations, one obtains
\begin{align*}
[J_x,J_y]&=2iJ_z,\\
[J_y,J_z]&=2iJ_x,\\
[J_z,J_x]&=2iJ_y.
\end{align*}
Equivalently,
\begin{equation*}
[J_x,J_z]=-2iJ_y,
\end{equation*}
which gives
\begin{equation*}
J_y=\frac{i}{2}[J_x,J_z].
\end{equation*}

The LEO Hamiltonian is
\begin{equation*}
H_C=\mu(t)\left(c_0^\dagger c_0+c_1^\dagger c_1\right).
\end{equation*}
Since \(H_C\) is proportional to the total particle-number operator, it
commutes with all single-IPQ generators,
\begin{equation*}
[H_C,J_x]=[H_C,J_y]=[H_C,J_z]=0.
\end{equation*}
It also commutes with the C-phase interaction,
\begin{equation*}
[H_C,H_{\rm cp}]=0.
\end{equation*}

\SMstart
\section{The binomial-code example}
\label{app:binomial}

This Appendix provides the details of the representative binomial-code
example discussed in Sec.~\ref{sec:binomial_comparison}. Consider a code
space spanned by the logical states
\(\{|W_0\rangle,|W_1\rangle\}\) and an error set \(\{E_k\}\).
The Knill--Laflamme condition requires
\begin{equation}
\langle W_i|E_l^\dagger E_k|W_j\rangle
=
C_{lk}\delta_{ij},
\label{eq:KL_binomial}
\end{equation}
where \(C_{lk}\) is independent of the logical indices \(i\) and \(j\).

For a representative code that corrects a single photon-loss event, the
logical codewords are
\begin{equation}
|W_0\rangle
=
\frac{|0\rangle+|4\rangle}{\sqrt{2}},
\qquad
|W_1\rangle
=
|2\rangle,
\label{eq:binomial_codewords}
\end{equation}
where \(|n\rangle\) denotes a Fock state of a single bosonic mode. An
arbitrary encoded state is
\begin{equation}
|\psi\rangle
=
u|W_0\rangle+v|W_1\rangle,
\qquad
|u|^2+|v|^2=1.
\end{equation}

For the single-photon-loss operator \(\hat a\), one obtains
\begin{equation}
\hat a|W_0\rangle
=
\sqrt{2}\,|3\rangle,
\qquad
\hat a|W_1\rangle
=
\sqrt{2}\,|1\rangle.
\label{eq:binomial_loss}
\end{equation}
The two error states are orthogonal and have the same norm, so that
\begin{equation}
\langle W_i|\hat a^\dagger\hat a|W_j\rangle
=
2\delta_{ij}.
\end{equation}
Thus, the Knill--Laflamme condition is satisfied for the
single-photon-loss error.

The photon-number parity operator is
\begin{equation}
P_B
=
e^{i\pi\hat n},
\qquad
P_B|n\rangle
=
(-1)^n|n\rangle,
\end{equation}
where \(\hat n=\hat a^\dagger\hat a\). The original code space has even
parity, whereas a single-photon-loss event maps it to the odd-parity
subspace spanned by \(|3\rangle\) and \(|1\rangle\).

Conditioned on the detection of a photon-loss event, a recovery unitary can
be defined by
\begin{equation}
U_{\rm rec}|3\rangle
=
|W_0\rangle,
\qquad
U_{\rm rec}|1\rangle
=
|W_1\rangle.
\end{equation}
It follows that
\begin{equation}
U_{\rm rec}\hat a|\psi\rangle
=
\sqrt{2}
\left(
u|W_0\rangle+v|W_1\rangle
\right),
\end{equation}
so that the original logical state is restored after normalization.

This recovery differs from the IPQ procedure. The binomial code satisfies the
Knill--Laflamme condition and permits a conditional unitary recovery, whereas
the first-order IPQ error \(x=x_0+x_1\) is corrected through the
measurement-assisted inverse operation described in Sec.~II\,B.
\SMend

\section{Collective decoherence: derivations}\label{app:collective}

\subsection{Dynamics in the encoding subspace (collective reservoir)}
We encode the information in the operator subspace spanned by \(\{c_{i}^{0}, c_{i}^{0\dagger}\}\).
Decoherence causes information to leak from this subspace, leading to failures in both information
storage and gate operations. First, we examine the dynamics of the encoding subspace in the \(a_{i}\)
representation. Specifically, we focus on the coefficients \(A_{i}^{0}\) and \(A_{i}^{1}\), whose evolution
is governed by the following equations:
\begin{eqnarray}
\dot{A}_{0}^{0}(t) & = & -i\,\left(\mu(t)-G_{x}\right)A_{0}^{0}(t)-i\,G_{z}A_{0}^{1}(t),\label{eq:c00}\\
\dot{A}_{1}^{0}(t) & = & -i\,\left(\mu(t)-G_{x}\right)A_{1}^{0}(t)-i\,G_{z}A_{1}^{1}(t)\label{eq:c01}\\
\dot{A}_{0}^{1}(t) & = & -i\,\left(\mu(t)+G_{x}\right)A_{0}^{1}(t)-i\,G_{z}A_{0}^{0}(t)-\int_{0}^{t}\text{d}s\,f\left(t-s\right)A_{0}^{1}(s),\label{eq:c10}\\
\dot{A}_{1}^{1}(t) & = & -i\,\left(\mu(t)+G_{x}\right)A_{1}^{1}(t)-i\,G_{z}A_{1}^{0}(t)-\int_{0}^{t}\text{d}s\,f\left(t-s\right)A_{1}^{1}(s).\label{eq:c11}
\end{eqnarray}
In what follows, we consider three different gate operations: information
storage, X-gate operation and Z-gate operation.

\subsubsection{Information Storage.}

In this case, no gate operation is applied, i.e., \(G_{x} = G_{z} = 0\). The information is encoded in
the operator subspace with the basis \(\{a_{i}^{0}\}\) at the initial moment, meaning
\(A_{0}^{0}(0) = A_{1}^{1}(0) = 1\) and \(A_{0}^{1}(0) = A_{1}^{0}(0) = 0\). At this point,
the evolutions of \(A_{i}^{j}(t)\) are decoupled from each other:
\begin{eqnarray*}
\dot{A}_{0}^{0}(t) & = & -i\,\mu(t)A_{0}^{0}(t),\\
\dot{A}_{1}^{0}(t) & = & -i\,\mu(t)A_{1}^{0}(t),\\
\dot{A}_{0}^{1}(t) & = & -i\,\mu(t)A_{0}^{1}(t)-\int_{0}^{t}\text{d}s\,f\left(t-s\right)A_{0}^{1}(s),\\
\dot{A}_{1}^{1}(t) & = & -i\,\mu(t)A_{1}^{1}(t)-\int_{0}^{t}\text{d}s\,f\left(t-s\right)A_{1}^{1}(s).
\end{eqnarray*}
Introducing slow variants $\tilde{A}_{i}^{j}(t)=A_{i}^{j}(t)\exp(-i\,\int_{0}^{t}\text{d}\tau\,\mu(\tau)),$
we have
\[
\dot{\tilde{A_{i}}}^{0}(t)=0,
\]
\begin{eqnarray*}
\dot{\tilde{A_{i}}}^{1}(t) & = & -\exp\left(-i\,\int_{0}^{t}\text{d}\tau\,\mu(\tau)\right)\int_{0}^{t}\text{d}s\,f\left(t-s\right)\exp\left(i\,\int_{0}^{s}\text{d}\tau\,\mu(\tau)\right)\tilde{A}_{i}^{j}(s).
\end{eqnarray*}
 As a result, the information encoded in the operator subspace is
robust to against collective noise at a finite reservoir temperature. We
can rewrite it into a vector equation as
\begin{eqnarray}
\dot{\tilde{\vec{A}}}(t) & = & -\exp\left(-i\,\int_{0}^{t}\text{d}\tau\,\mu(\tau)\right)\int_{0}^{t}\text{d}s\,\bar{f}\left(t-s\right)\exp\left(i\,\int_{0}^{s}\text{d}\tau\,\mu(\tau)\right)\tilde{\vec{A}}(s)\label{eq:is}
\end{eqnarray}
with $\tilde{\vec{A}}(t)=[\tilde{A}_{1}^{1}(t),\tilde{A}_{0}^{1}(t),\tilde{A}_{1}^{0}(t),\tilde{A}_{0}^{0}(t)],$
and
\[
\bar{f}=\text{diag}(f,f,0,0).
\]

\subsubsection{The X-Gate operation.}

When we consider a X-gate operation on the IPQ, it requires $G_{z}=0$, which leads to
\begin{eqnarray*}
\dot{A}_{0}^{0}(t) & = & -i\,\left(\mu(t)-G_{x}\right)A_{0}^{0}(t),\\
\dot{A}_{1}^{0}(t) & = & -i\,\left(\mu(t)-G_{x}\right)A_{1}^{0}(t),\\
\dot{A}_{0}^{1}(t) & = & -i\,\left(\mu(t)+G_{x}\right)A_{0}^{1}(t)-\int_{0}^{t}\text{d}s\,f\left(t-s\right)A_{0}^{1}(s),\\
\dot{A}_{1}^{1}(t) & = & -i\,\left(\mu(t)+G_{x}\right)A_{1}^{1}(t)-\int_{0}^{t}\text{d}s\,f\left(t-s\right)A_{1}^{1}(s).
\end{eqnarray*}
Similar to information storage, the information is encoded in the encoding subspace with the basis \(\{a_{i}^{0}\}\).
We introduce \(\tilde{A}_{i}^{j}(t)\), which has the same definition as before. We have
\begin{eqnarray*}
\dot{\tilde{A_{i}^{0}}}(t) & = & i\,G_{x}\tilde{A}_{i}^{0}(t),\\
\dot{\tilde{A_{i}^{1}}}(t) & = & -i\,G_{x}\tilde{A}_{i}^{1}(t)-\exp\left(-i\,\int_{0}^{t}\text{d}\tau\,\mu(\tau)\right)\int_{0}^{t}\text{d}s\,f\left(t-s\right)\exp\left(i\,\int_{0}^{s}\text{d}\tau\,\mu(\tau)\right)\tilde{A}_{i}^{1}(s),
\end{eqnarray*}
which can be written as a vector form
\begin{eqnarray*}
\dot{\tilde{\vec{A}}}(t) & = & -i\,G_{x}\bar{\sigma}_{z}\otimes\bar{I}_{2}{\tilde{\vec{A}}}(t)-\exp\left(-i\,\int_{0}^{t}\text{d}\tau\,\mu(\tau)\right)\int_{0}^{t}\text{d}s\,\exp\left(i\,\int_{0}^{s}\text{d}\tau\,\mu(\tau)\right)\bar{f}\left(t-s\right)\tilde{\vec{A}}(s),
\end{eqnarray*}
with same variants used in Eq.(\ref{eq:is}). Due to rapid oscillation
terms in the convolution, the second term in above equations
have no contribution on the X-gate operation. Therefore, the effective
dynamics will be described by
\[
\dot{\tilde{\vec{A}}}(t)=-i\,G_{x}\bar{\sigma}_{z}\otimes\bar{I}_{2}\tilde{\vec{A}}(t)
\]
 with $\tilde{\vec{A}}(t)=[\tilde{A}_{1}^{1}(t),\tilde{A}_{0}^{1}(t),\tilde{A}_{1}^{0}(t),\tilde{A}_{0}^{0}(t)]$,
which is perfect X-gate operation on the IPQ without the decoherence
effect.

\subsubsection{The Z-Gate Operation.}

Finally, we consider the Z-gate operation with \(G_{x} = 0\). As the information needs to
be transferred between \(a_{0}^{0}\) and \(a_{1}^{0}\), all expansion coefficients of \(a_{0}^{0}\)
and \(a_{1}^{0}\) must be accounted for, with their dynamics described by Eqs.
(\ref{eq:c00} - \ref{eq:c11}). Then we can write down
\begin{eqnarray*}
 &  & \dot{A}_{0}^{0}(t)=-i\,\mu(t)A_{0}^{0}(t)-i\,G_{z}A_{0}^{1}(t),\\
 &  & \dot{A}_{1}^{0}(t)=-i\,\mu(t)A_{1}^{0}(t)-i\,G_{z}A_{1}^{1}(t),\\
 &  & \dot{A}_{0}^{1}(t)=-i\,\mu(t)A_{0}^{1}(t)-i\,G_{z}A_{0}^{0}(t)-\int_{0}^{t}\text{d}s\,f\left(t-s\right)A_{0}^{1}(s),\\
 &  & \dot{A}_{1}^{1}(t)=-i\,\mu(t)A_{1}^{1}(t)-i\,G_{z}A_{1}^{0}(t)-\int_{0}^{t}\text{d}s\,f\left(t-s\right)A_{1}^{1}(s).
\end{eqnarray*}
We can rewrite the equations into a vector equation,
\[
i\dot \vec{A}(t)=\left(\mu(t)\bar{I}_{4}+G_{z}\bar{\sigma}_{x}\otimes\bar{I_{2}}\right)\vec{A}(t)+\int_{0}^{t}\text{d}s\,\bar{f}\left(t-s\right)\vec{A}(s).
\]
 By using unitary matrix
\[
\bar{U}_{Z}=\frac{\sqrt{2}}{2}\left(\begin{array}{cccc}
1 & 0 & -1 & 0\\
0 & 1 & 0 & -1\\
1 & 0 & 1 & 0\\
0 & 1 & 0 & 1
\end{array}\right),
\]
 the vector equation can be transformed into
\[
\dot{\vec{D}}(t)=-i\,\left(\mu(t)\bar{I}_{4}+G_{z}\bar{\sigma}_{z}\otimes\bar{I_{2}}\right)\vec{D}(t)-\int_{0}^{t}\text{d}s\,\bar{F}\left(t-s\right)\vec{D}(s)
\]
with $\vec{D}(t)=\bar{U}_{Z}\vec{A}$ and
\[
\bar{F}=\bar{U}_{Z}\bar f \bar{U}_{Z}^\dagger=\frac{1}{2}\left(\begin{array}{cccc}
f & 0 & -f & 0\\
0 & f & 0 & -f\\
-f & 0 & f & 0\\
0 & -f & 0 & f
\end{array}\right)
\]
By defining $\tilde{\vec{D}}(t)=\vec{D}(t)\exp(-i\,\int_{0}^{t}\text{d}\tau\,\mu(\tau))$,
we arrive at
\begin{eqnarray*}
\dot{\tilde{\vec{D}}}(t) & = & -i\,G_{z}\bar{\sigma}_{z}\otimes\bar{I_{2}}\tilde{\vec{D}}(t)\\
 &  & -\exp\left(-i\,\int_{0}^{t}\text{d}\tau\,\mu(\tau)\right)\int_{0}^{t}\text{d}s\,
 \exp\left(i\,\int_{0}^{s}\text{d}\tau\,\mu(\tau)\right)\bar{F}\left(t-s\right)\tilde{\vec{D}}(s).
\end{eqnarray*}
 Due to rapid oscillation caused by $\exp\left(\pm i\,\int_{0}^{s}\text{d}\tau\,\mu(\tau)\right)$,
the last term of $\dot{\tilde{\vec{D}}}(t)$ are canceled
in the evolution. Thus we have a more concise effective form:
\[
\dot{\vec{D}}(t)=-i\,G_{z}\bar{\sigma}_{z}\otimes\bar{I_{2}}\vec{D}(t).
\]
When we transform variants $\vec{D}$ back into $\vec{A}=U_{D}^{\dagger}\vec{D}$,
we obtain a perfect Z-gate operation as
\[
\dot\vec{A}(t)=-i\,G_{z}\bar{\sigma}_{x}\otimes\bar{I_{2}}\vec{A}(t).
\]

The coeffcient vector can be tranformed form the $\{ {a}_{i}\}$
representation to the $\{ {c}_{i}\}$ representation by means of
a constant unitary matrix:
\[
\bar{U}_{A}=\frac{1}{2}\left(\begin{array}{cccc}
1 & 1 & 1 & 1\\
1 & 1 & -1 & -1\\
1 & -1 & 1 & -1\\
1 & -1 & -1 & 1
\end{array}\right),
\]
which leads to $\vec{C}(t)=\bar{U}_{A}\vec{A}(t)$.  The gate operation matrices in the
\(\{c_{i}\}\) representation satisfy
\(\bar{\sigma}_{x} \otimes \bar{I}_{2} = \bar{U}_{A} \bar{\sigma}_{z} \otimes \bar{I}_{2} \bar{U}_{A}^{\dagger}\)
and \(\bar{\sigma}_{z} \otimes \bar{I}_{2} = \bar{U}_{A} \bar{\sigma}_{x} \otimes \bar{I}_{2} \bar{U}_{A}^{\dagger}\).
The correlation function matrix is given by \(\bar{F}_{C}(t-s) = \bar{U}_{A} \bar{F}(t-s) \bar{U}_{A}^{\dagger}\).
 Since \(\bar{U}_{A}\) is a constant matrix, the action of LEO pulses remains effective. As a result, we can obtain
 the dynamical equations for the encoding subspace in the \(\{c_{i}\}\) representation.

\subsection{Gate-operation dynamics}
To characterize the decoherence dynamics of a single identical particle, we examine the time evolution
of the average excitation number $\langle {n}_{i}\rangle$ for the $i$th identical particle. We assume
that, initially, the system and the heat reservoirs are uncorrelated, with the total density matrix expressed
as $\rho_{\text{tot}}=\rho_{\text{S}}(0)\otimes\rho_{\text{B}}(0)$. The reservoir is prepared in a thermal
equilibrium state:
\[
\rho_{\text{B}}(0)=\sum_{n,\alpha}\frac{e^{-\beta\omega_{\alpha}}}{1-e^{-\beta\omega_{\alpha}}}\left|n_{\alpha}\right\rangle \left\langle n_{\alpha}\right|,
\]
where $\beta$ is the inverse reservoir temperature, and $\left|n_{\alpha}\right\rangle$ is the Fock state
with excitation number $n_{\alpha}$. Thus, we have
\[
\langle {n}_{i}(t)\rangle = \langle {c}_{i}^{\dagger}(t) {c}_{i}(t)\rangle = \sum_{j,k=0,1}C_{j}^{i*}(t)C_{k}^{i}(t)\langle {c}_{j}^{0\dagger} {c}_{k}^{0}\rangle_{\text{S}} + \sum_{\alpha}|B_{c,\alpha}^{i}(t)|^{2}\langle {b}_{\alpha}^{0\dagger} {b}_{\alpha}^{0}\rangle_{\text{B}},
\]
where $\text{Tr}_{B}\{ {b}_{\beta}^{0\dagger} {b}_{\alpha}^{0}\rho_{\text{B}}(0)\}=\delta_{\alpha\beta}\bar n(\omega_\alpha)$
has been taken into account.

 \subsubsection{The X-gate operation.}

In what follows, we simulate the average values of the observable variables
\(\langle J_{x}(t) \rangle\), \(\langle J_{y}(t) \rangle\), and \(\langle J_{z}(t) \rangle\),
where \(\langle J_{k}(t) \rangle = \text{Tr}\{J_{k}(t)\rho_{\text{tot}}\}\).
We consider the expected values of the "Pauli operators" during the X-gate operation.
The x-component is
\begin{eqnarray*}
\left\langle J_{x}(t)\right\rangle  & = & \left\langle  {a}_{1}^{\dagger}(t) {a}_{1}(t)- {a}_{0}^{\dagger}(t) {a}_{0}(t)\right\rangle \\
 & = & \sum_{i,j=0,1}\left(A_{j}^{1*}(t)A_{i}^{1}(t)-A_{j}^{0*}(t)A_{i}^{0}(t)\right)\left\langle  {a}_{j}^{0\dagger} {a}_{i}^{0}\right\rangle _{\text{S}}\\
 &  & +\sum_{\alpha,\beta}\left(B_{\beta}^{1*}(t)B_{\alpha}^{1}(t)-B_{\beta}^{0*}(t)B_{\alpha}^{0}(t)\right)\left\langle  {b}_{\beta}^{0\dagger} {b}_{\alpha}^{0}\right\rangle _{\text{B}}\\
 & = & \sum_{i,j=0,1}\left(A_{j}^{1*}(t)A_{i}^{1}(t)-A_{j}^{0*}(t)A_{i}^{0}(t)\right)\left\langle  {a}_{j}^{0\dagger} {a}_{i}^{0}\right\rangle _{\text{S}}\\
 &  & +\sum_{\alpha}\left(|B_{\alpha}^{1}(t)|^{2}-|B_{\alpha}^{0}(t)|^{2}\right)\left\langle  {b}_{\alpha}^{0\dagger} {b}_{\alpha}^{0}\right\rangle _{\text{B}}.
\end{eqnarray*}
We can calculate $B_{\beta}^{i}$ by means of the
Laplace transform. Since $ {a}_{0}(t)$ decouples to
the reservoir, we have $B_{\alpha}^{0}(t)=0$.
 The remaining coefficient obeys
\begin{equation}
\dot{B}_{\alpha}^{1}(t) = -i\,\left(\mu+G_{x}\right)B_{\alpha}^{1}(t)-\int_{0}^{t}\text{d}s\,f\left(t-s\right)B_{\alpha}^{1}(s)
-i\,g_{\alpha}\exp\left(-i\,\omega_{\alpha}t\right),
\label{eq:dbi1}
\end{equation}
where \(f(t-s)\) is the bath correlation function. Performing the Laplace transformation of Eq.~(\ref{eq:dbi1}), we have
\begin{eqnarray*}
\tilde{B}_{\alpha}^{1}(t_{1}) & = & \frac{-i\,g_{\alpha}}{\left(t_{1}+i\,\omega_{\alpha}\right)\left(t_{1}+\tilde{f}\left(t_{1}\right)+i\,\left(\mu+G_{x}\right)\right)},
\end{eqnarray*}
where $\tilde{f}\left(t_{1}\right)$ are the Laplace transformation
of the correlation function for the reservoir, and $B_{\alpha}^{1}(0)=0$
are considered. If the correlation function $f(t-s)$ is given, the
solution of can be obtained by taking the inverse Laplace transformation
on $\tilde{B}_{\alpha}^{1}(t_{1})$, i.e., $B_{\beta}^{1}(t)=\mathcal{L}^{-1}[\tilde{B}_{\alpha}^{1}(t_{1})]$.
By introducing
\begin{eqnarray*}
\tilde{B'}_{\alpha}^{1}(t_{1}) & = & \frac{1}{\left(t_{1}+i\,\omega_{\alpha}\right)\left(t_{1}+\tilde{f}\left(t_{1}\right)+i\,\left(\mu+G_{x}\right)\right)},
\end{eqnarray*}
we can see that $B_{\beta}^{1}(t)=-ig_{\alpha}B'{}_{\beta}^{1}(t)$
with $B'{}_{\beta}^{1}(t)=\mathcal{L}^{-1}[\tilde{B'}_{\alpha}^{1}(t_{1})]$,
which leads to
\begin{eqnarray*}
\langle J_{x}(t)\rangle & = & \sum_{i,j=0,1}\left(A_{j}^{1*}(t)A_{i}^{1}(t)-A_{j}^{0*}(t)A_{i}^{0}(t)\right)\langle {a}_{j}^{0\dagger} {a}_{i}^{0}\rangle_{\text{S}}\\
 &  & +\sum_{i,\alpha}|g_{\alpha}|^{2}|B'{}_{\alpha}^{1}(t)|^{2}\langle {b}_{\alpha}^{0\dagger} {b}_{\alpha}^{0}\rangle_{\text{B}}\\
 & = & \sum_{i,j=0,1}\left(A_{j}^{1*}(t)A_{i}^{1}(t)-A_{j}^{0*}(t)A_{i}^{0}(t)\right)\langle {a}_{j}^{0\dagger} {a}_{i}^{0}\rangle_{\text{S}}\\
 &  & +\sum_{\alpha}|g_{\alpha}|^{2}|B'{}_{\alpha}^{1}(t)|^{2}\frac{1}{e^{\beta\omega_{\alpha}}-1}\\
 & = & \sum_{i,j=0,1}\left(A_{j}^{1*}(t)A_{i}^{1}(t)-A_{j}^{0*}(t)A_{i}^{0}(t)\right)\langle {a}_{j}^{0\dagger} {a}_{i}^{0}\rangle_{\text{S}}\\
 &  & +\int d\omega J(\omega)\mathcal B_{1}(\omega)\frac{1}{e^{\beta\omega}-1}.
\end{eqnarray*}
Here, we have considered the continuous limit of the reservoir modes,
i.e., $\sum_{\alpha}|g_{\alpha}|^{2}|B'{}_{\alpha}^{1}(t)|^{2}\rightarrow\int d\omega J(\omega)\mathcal B_{1}(\omega).$
With a similar procedure, $\langle J_{z}(t)\rangle$and $\langle J_{y}(t)\rangle$
are given by
\begin{eqnarray*}
\langle J_{z}(t)\rangle & = & \sum_{i,j=0,1}\left(A_{i}^{1*}(t)A_{j}^{0}(t)+A_{i}^{0*}(t)A_{j}^{1}(t)\right)\left\langle  {a}_{i}^{0\dagger} {a}_{j}^{0}\right\rangle _{\text{S}},\\
\langle J_{y}(t)\rangle & = & \sum_{i,j=0,1}i\,\left(A_{i}^{1*}(t)A_{j}^{0}(t)-A_{i}^{0*}(t)A_{j}^{1}(t)\right)\left\langle  {a}_{i}^{0\dagger} {a}_{j}^{0}\right\rangle _{\text{S}},
\end{eqnarray*}
 where $B_{\alpha}^{0}(t)=0$ and $\left\langle  {b}_{\beta}^{0\dagger} {b}_{\alpha}^{0}\right\rangle _{\text{B}}=\delta_{\alpha\beta}\frac{1}{e^{\beta\omega_{\alpha}}-1}$
have been used.

\subsubsection{The Z-gate operation.}

The dynamical equations of $B_{\alpha}^{i}(t)$  read
\begin{eqnarray*}
\dot{B}_{\alpha}^{0}(t) & = & -i\,\mu B_{\alpha}^{0}(t)-i\,G_{z}B_{\alpha}^{1}(t),\\
\dot{B}_{\alpha}^{1}(t) & = & -i\,\mu B_{\alpha}^{1}(t)-i\,G_{z}B_{\alpha}^{0}(t)-\int_{0}^{t}\text{d}s\,f\left(t-s\right)B_{\alpha}^{1}(s),\\
 &  & -i\,g_{\alpha}\exp\left(-i\:\omega_{\alpha}t\right),
\end{eqnarray*}
We can solve it by the Laplace transformation, which leads to
\begin{eqnarray*}
\tilde{B}_{\alpha}^{0}(t_{1}) & = & \frac{g_{\alpha}\,G_{z}}{\left(t_{1}+i\,\omega_{\alpha}\right)\,\left(G_{z}^{2}+\left(t_{1}+i\,\mu\right)\,\left(t_{1}+i\,\mu+\tilde{f}(t_{1})\right)\right)},\\
\tilde{B}_{\alpha}^{1}(t_{1}) & = & \frac{g_{\alpha}\,\left(t_{1}+i\,\mu\right)}{\left(t_{1}+i\,\omega_{\alpha}\right)\,\left(G_{z}^{2}+\left(t_{1}+i\,\mu)\right)\,\left(t_{1}+i\,\mu+\tilde{f}(t_{1})\right)\right)}.
\end{eqnarray*}
where $\tilde{f}_{k}(t_{1})=\mathcal{L}[f_{k}(t)]$ is the result
of the Laplace tranformation about $f_{k}(t)$, and $B_{j}^{i}(0)=0$
is used. If $f_{k}(t)$ is given, we can obtain the solutions by means
of the inverse Laplace transformation. The mean values of $\{ {J}_{z}\}$
read
\begin{eqnarray*}
\left\langle J_{z}(t)\right\rangle  & = & \sum_{i,j=0,1}\left(A_{i}^{1*}(t)A_{j}^{0}(t)+A_{i}^{0*}(t)A_{j}^{1}(t)\right)\left\langle  {a}_{i}^{0\,\dagger} {a}_{j}^{0}\right\rangle _{\text{S}}\\
 & + & \sum_{i,\alpha}|g_{\alpha}|^{2}\left(B'{}_{\alpha}^{0*}(t)B_{\alpha}^{'1}(t)+B'{}_{\alpha}^{1*}(t)B'{}_{\alpha}^{0}(t)\right)\frac{1}{e^{\beta\omega_{\alpha}}-1},
\end{eqnarray*}
\begin{eqnarray*}
\left\langle J_{y}(t)\right\rangle  & = & \sum_{i,j=0,1}i\,\left(A_{i}^{1*}(t)A_{j}^{0}(t)-A_{i}^{0*}(t)A_{j}^{1}(t)\right)\left\langle  {a}_{i}^{0\,\dagger} {a}_{j}^{0}\right\rangle _{\text{S}}\\
 & + & \sum_{i,\alpha}i\,|g_{\alpha}|^{2}\left(B'{}_{\alpha}^{0*}(t)B'{}_{\alpha}^{1}(t)-B'{}_{\alpha}^{1*}(t)B'{}_{\alpha}^{0}(t)\right)\frac{1}{e^{\beta\omega_{\alpha}}-1},
\end{eqnarray*}
\begin{eqnarray*}
\left\langle J_{x}(t)\right\rangle  & = & \sum_{i,j=0,1}\left(A_{j}^{1*}(t)A_{i}^{1}(t)-A_{j}^{0*}(t)A_{i}^{0}(t)\right)\left\langle  {a}_{j}^{0\dagger} {a}_{i}^{0}\right\rangle _{\text{S}}\\
 &  & +\sum_{i,\alpha}|g_{\alpha}|^{2}\left(B'{}_{\alpha}^{1*}(t)B'{}_{\alpha}^{1}(t)-B'{}_{\alpha}^{0*}(t)B'{}_{\alpha}^{0}(t)\right)\frac{1}{e^{\beta\omega_{\alpha}}-1},
\end{eqnarray*}
with $\left\langle  {b}_{\beta}^{0\dagger} {b}_{\alpha}^{0}\right\rangle _{\text{B}}=\delta_{\alpha\beta}\frac{1}{e^{\beta\omega_{\alpha}}-1}$
.

\section{Individual decoherence: derivations}\label{app:individual}

The evolution of $ {c}_{i}(t)$ in the Heisenberg picture fulfills
\begin{eqnarray}
\dot{ {c}}_{0} & = & i\,\left[ {H}_{\text{tot}}, {c}_{0}\right]\nonumber \\
 & = &-i\,\left(\omega_{0,0}+\mu(t)-G_{z}\right) {c}_{0}-i\,G_{x} {c}_{1}-i\,\sum_{\alpha}g_{0,\alpha}\left( {b}_{0,\alpha}^{\dagger}+ {b}_{0,\alpha}\right),\label{eq:c0}
\end{eqnarray}
\begin{eqnarray}
\dot{ {c}}_{1} & = & i\,\left[ {H}_{\text{tot}}, {c}_{1}\right]\nonumber \\
 & = & -i\,\left(\omega_{0,1}+\mu(t)+G_{z}\right) {c}_{1}-i\,G_{x} {c}_{0}-i\,\sum_{\alpha}g_{1,\alpha}\left( {b}_{1,\alpha}^{\dagger}+ {b}_{1,\alpha}\right),\label{eq:c1}
\end{eqnarray}
where
\begin{eqnarray*}
\left[ {J}_{x}, {c}_{0}\right] & = & - {c}_{1},\:\left[ {J}_{z}, {c}_{0}\right]= {c}_{0},\\
\left[ {J}_{x}, {c}_{1}\right] & = & - {c}_{0},\:\left[ {J}_{z}, {c}_{1}\right]=- {c}_{1}.
\end{eqnarray*}
have been used. And the reservoir annihilation operator $ {b}_{i,\alpha}(t)$ satisfies
\begin{eqnarray}
\dot{ {b}}_{i,\alpha} & = & i\,\left[ {H}_{\text{tot}}, {b}_{i,\alpha}\right]\nonumber \\
 & = & -i\,\omega_{i,\alpha} {b}_{i,\alpha}-i\,g_{i,\alpha}\left( {c}_{i}^{\dagger}+ {c}_{i}\right).\label{eq:bi}
\end{eqnarray}
By formally integrating Eq. (\ref{eq:bi}), the equation of motion
for the reservoir operator is
\begin{eqnarray*}
 {b}_{i,\alpha}(t) & = &  {b}_{i,\alpha}(0)\exp\left(-i\:\omega_{i,\alpha}t\right)\\
 &  & -i\,g_{i,\alpha}\int_{0}^{t}\text{d}s\,\left( {c}_{i}^{\dagger}(s)+ {c}_{i}(s)\right)\exp\left(-i\:\omega_{i,\alpha}\left(t-s\right)\right).
\end{eqnarray*}
\begin{eqnarray*}
 {b}_{i,\alpha}^{\dagger}(t) & = &  {b}_{i,\alpha}^{\dagger}(0)\exp\left(i\:\omega_{i,\alpha}t\right)\\
 &  & +i\,g_{i,\alpha}\int_{0}^{t}\text{d}s\,\left( {c}_{i}^{\dagger}(s)+ {c}_{i}(s)\right)\exp\left(i\:\omega_{i,\alpha}\left(t-s\right)\right).
\end{eqnarray*}
Substituting the formal solution into Eqs. (\ref{eq:c0}) and (\ref{eq:c1}),
we have
\begin{eqnarray*}
\dot{ {c}}_{0}(t) & = & -i\,\left(\omega_{0,0}+\mu(t)-G_{z}\right) {c}_{0}(t)-i\,G_{x} {c}_{1}(t)\\
 &  & -i\,\sum_{\alpha}g_{0,\alpha}\left( {b}_{0,\alpha}^{\dagger}(0)\exp\left(i\:\omega_{0,\alpha}t\right)+ {b}_{0,\alpha}(0)\exp\left(-i\:\omega_{0,\alpha}t\right)\right)\\
 &  & -2i\int_{0}^{t}\text{d}s\,\left( {c}_{0}^{\dagger}(s)+ {c}_{0}(s)\right)\sum_{\alpha}g_{0,\alpha}^{2}\sin\left(\omega_{0,\alpha}\left(t-s\right)\right)
\end{eqnarray*}
and
\begin{eqnarray*}
\dot{ {c}}_{1}(t)  & = & -i\,\left(\omega_{0,1}+\mu(t)+G_{z}\right) {c}_{1}(t)-i\,G_{x} {c}_{0}(t)\\
 &  & -i\,\sum_{\alpha}g_{1,\alpha}\left( {b}_{1,\alpha}(0)\exp\left(-i\:\omega_{1,\alpha}t\right)+ {b}_{1\alpha}^{\dagger}(0)\exp\left(i\:\omega_{1,\alpha}t\right)\right)\\
 &  & -2i\int_{0}^{t}\text{d}s\,\left( {c}_{1}^{\dagger}(s)+ {c}_{1}(s)\right)\sum_{\alpha}g_{1,\alpha}^{2}\sin\left(\omega_{1,\alpha}\left(t-s\right)\right)
\end{eqnarray*}

Inserting the linear expansion of the system operators used in the main text into the equations below yields the coefficient equations for $C_i^j(t)$, $\bar{C}_i^j(t)$, $B_{i,\alpha}^j(t)$, and $\bar{B}_{i,\alpha}^j(t)$.

\subsection{Dynamics in the encoding subspace (individual reservoirs)}
We will focus on components in the encoding subspace. We are interested
in the coefficients \( C_{i}^{0} \) and \( C_{i}^{1} \), whose evolution satisfies
\begin{eqnarray}
\dot{C}_{0}^{0}(t) & = & -i\,\left(\omega_{0,0}+\mu(t)-G_{z}\right)C_{0}^{0}(t)-i\,G_{x}C_{0}^{1}(t)-\int_{0}^{t}\text{d}s\,f_{0}\left(t-s\right)\left(\bar{C}_{0}^{0\,*}(s)+C_{0}^{0}(s)\right),\label{eq:c00-1}\\
\dot{C}_{1}^{0}(t) & = & -i\,\left(\omega_{0,0}+\mu(t)-G_{z}\right)C_{1}^{0}(t)-i\,G_{x}C_{1}^{1}(t)-\int_{0}^{t}\text{d}s\,f_{0}\left(t-s\right)\left(\bar{C}_{1}^{0\,*}(s)+C_{1}^{0}(s)\right),\label{eq:c01-1}\\
\dot{C}_{0}^{1}(t) & = & -i\,\left(\omega_{0,1}+\mu(t)+G_{z}\right)C_{0}^{1}(t)-i\,G_{x}C_{0}^{0}(t)-\int_{0}^{t}\text{d}s\,f_{1}\left(t-s\right)\left(\bar{C}_{0}^{1\,*}(s)+C_{0}^{1}(s)\right),\label{eq:c10-1}\\
\dot{C}_{1}^{1}(t) & = & -i\,\left(\omega_{0,1}+\mu(t)+G_{z}\right)C_{1}^{1}(t)-i\,G_{x}C_{1}^{0}(t)-\int_{0}^{t}\text{d}s\,f_{1}\left(t-s\right)\left(\bar{C}_{1}^{1\,*}(s)+C_{1}^{1}(s)\right).\label{eq:c11-1}
\end{eqnarray}
To apply gate operations on the identical particle qubits (IPQs), we set \( \omega_{0,0} = \omega_{0,1} = 0 \).
In the following sections, we will consider
three different gate operations: information storage, the X-gate operation, and the Z-gate operation.

{\subsubsection{Information storage process.} }
In this case, the gate operations are not applied, i.e., \( G_{x} = G_{z} = 0 \). The information is encoded in
the operator subspace with the basis \(\{ {c}_{i}^{0}\}\) at the initial moment, which implies
\( C_{0}^{0}(0) = C_{1}^{1}(0) = 1 \). At this time, the evolutions of \( C_{i}^{j}(t) \) are decoupled
from each other:
\begin{eqnarray*}
\dot{C}_{0}^{0}(t) & = & -i\,\mu(t)C_{0}^{0}(t)-\int_{0}^{t}\text{d}s\,f_{0}\left(t-s\right)C_{0}^{0}(s)+c_{0}^{0}(t), \\
\dot{C}_{1}^{0}(t) & = & -i\,\mu(t)C_{1}^{0}(t)-\int_{0}^{t}\text{d}s\,f_{0}\left(t-s\right)C_{1}^{0}(s)+c_{1}^{0}(t), \\
\dot{C}_{0}^{1}(t) & = & -i\,\mu(t)C_{0}^{1}(t)-\int_{0}^{t}\text{d}s\,f_{1}\left(t-s\right)C_{0}^{1}(s)+c_{0}^{1}(t), \\
\dot{C}_{1}^{1}(t) & = & -i\,\mu(t)C_{1}^{1}(t)-\int_{0}^{t}\text{d}s\,f_{1}\left(t-s\right)C_{1}^{1}(s)+c_{1}^{1}(t),
\end{eqnarray*}
where \( c_{i}^{j}(t) = \int_{0}^{t}\text{d}s\,f_{0}\left(t-s\right)\bar{C}_{i}^{j\,*}(s) \).
Introducing slow variables \( \tilde{C}_{i}^{j}(t) = C_{i}^{j}(t)\exp\left(-i\,\int_{0}^{t}\text{d}\tau\,\mu(\tau)\right) \),
we obtain
\begin{eqnarray*}
\dot{\tilde{C_{i}}}^{j}(t) & = & -\exp\left(-i\,\int_{0}^{t}\text{d}\tau\,\mu(\tau)\right) \\
 &  & \times\int_{0}^{t}\text{d}s\,f_{i}\left(t-s\right)\exp\left(i\,\int_{0}^{s}\text{d}\tau\,\mu(\tau)\right)\tilde{C}_{i}^{j}(s)+\tilde{c}_{i}^{j}(t),
\end{eqnarray*}
where \( \tilde{c}_{i}^{j}(t) = -i\int_{0}^{t}\text{d}s\,f_{0}\left(t-s\right)\bar{C}_{i}^{j\,*}(s)\exp\left(-i\,\int_{0}^{t}\text{d}\tau\,\mu(\tau)\right) \).

To address the rapidly oscillating behavior in \( \tilde{c}_{i}^{j}(t) \), we focus on the dynamical
equation for \( \bar{C}_{i}^{j\,*} \), given by
\[
\dot{\bar{C}}_{i}^{j\,*}(t) = i\,\mu(t)\bar{C}_{i}^{j\,*}(t) - \int_{0}^{t}\text{d}s\,f_{0}^{*}\left(t-s\right)\left(C_{i}^{j}(s) + \bar{C}_{i}^{j\,*}(s)\right).
\]
It appears that \( \bar{C}_{i}^{j\,*}(s) \) contains oscillating terms such as
\( \exp\left(i\,\int_{0}^{s}\text{d}\tau\,\mu(\tau)\right) \), leading to \( \tilde{c}_{i}^{j}(t) \to 0 \).
As a result, the information encoded in the operator subspace is robust against individual noise at finite
reservoir temperature. We can express this as a vector equation
\begin{eqnarray}
\dot{\tilde{\vec{C}}}(t) & = & -\exp\left(-i\,\int_{0}^{t}\text{d}\tau\,\mu(\tau)\right) \nonumber \\
 &  & \times \int_{0}^{t}\text{d}s\,\bar{f}\left(t-s\right)\exp\left(i\,\int_{0}^{s}\text{d}\tau\,\mu(\tau)\right)\tilde{\vec{C}}(s)+\tilde{\vec{c}}(t), \label{eq:is-1}
\end{eqnarray}
with \( \vec{C}(t) = [\tilde{C}_{1}^{1}(t), \tilde{C}_{0}^{1}(t), \tilde{C}_{1}^{0}(t), \tilde{C}_{0}^{0}(t)] \), \( \vec{c}(t) = [\tilde{c}_{1}^{1}(t), \tilde{c}_{0}^{1}(t), \tilde{c}_{1}^{0}(t), \tilde{c}_{0}^{0}(t)] \), and
\[
\bar{f} = \text{diag}(f_{1}, f_{1}, f_{0}, f_{0}).
\]

\subsubsection{The Z-Gate Operation.}

When we consider Z-gate operation on the IPQ, it requires \( G_{x} = 0 \), leading to the following equations
\begin{eqnarray*}
\dot{C}_{0}^{0}(t) & = & -i\,\left(\mu(t) - G_{z}\right)C_{0}^{0}(t) - \int_{0}^{t}\text{d}s\,f_{0}\left(t-s\right)\left(\bar{C}_{0}^{0\,*}(s) + C_{0}^{0}(s)\right), \\
\dot{C}_{1}^{0}(t) & = & -i\,\left(\mu(t) - G_{z}\right)C_{1}^{0}(t) - \int_{0}^{t}\text{d}s\,f_{0}\left(t-s\right)\left(\bar{C}_{1}^{0\,*}(s) + C_{1}^{0}(s)\right), \\
\dot{C}_{0}^{1}(t) & = & -i\,\left(\mu(t) + G_{z}\right)C_{0}^{1}(t) - \int_{0}^{t}\text{d}s\,f_{1}\left(t-s\right)\left(\bar{C}_{0}^{1\,*}(s) + C_{0}^{1}(s)\right), \\
\dot{C}_{1}^{1}(t) & = & -i\,\left(\mu(t) + G_{z}\right)C_{1}^{1}(t) - \int_{0}^{t}\text{d}s\,f_{1}\left(t-s\right)\left(\bar{C}_{1}^{1\,*}(s) + C_{1}^{1}(s)\right).
\end{eqnarray*}
Similar to information storage, the information is encoded in the operator subspace with the basis \(\{ {c}_{i}^{0}\}\).
We introduce \( \tilde{C}_{i}^{j}(t) \) with the same definition as before. Therefore, we have
\begin{eqnarray*}
\dot{\tilde{C}_{i}^{0}}(t) & = & i\,G_{z}\tilde{C}_{i}^{0}(t) - \exp\left(-i\,\int_{0}^{t}\text{d}\tau\,\mu(\tau)\right)
 \int_{0}^{t}\text{d}s\,f_{0}\left(t-s\right)\exp\left(i\,\int_{0}^{s}\text{d}\tau\,\mu(\tau)\right)\tilde{C}_{i}^{0}(s) + \tilde{c}_{i}^{0}(t), \\
\dot{\tilde{C}_{i}^{1}}(t) & = & -i\,G_{z}\tilde{C}_{i}^{1}(t) - \exp\left(-i\,\int_{0}^{t}\text{d}\tau\,\mu(\tau)\right)
  \int_{0}^{t}\text{d}s\,f_{1}\left(t-s\right)\exp\left(i\,\int_{0}^{s}\text{d}\tau\,\mu(\tau)\right)\tilde{C}_{i}^{1}(s) + \tilde{c}_{i}^{0}(t).
\end{eqnarray*}
These can be written in vector form as
\begin{eqnarray*}
\dot{\tilde{\vec{C}}}(t) & = & -i\,G_{z}\bar{\sigma}_{z} \otimes \bar{I}_{2} \dot{\tilde{\vec{C}}}(t) - \exp\left(-i\,\int_{0}^{t}\text{d}\tau\,\mu(\tau)\right) \int_{0}^{t}\text{d}s\,\exp\left(i\,\int_{0}^{s}\text{d}\tau\,\mu(\tau)\right)\bar{f}\left(t-s\right)\tilde{\vec{C}}(s) + \tilde{\vec{c}}(t),
\end{eqnarray*}
with the same variables used in Eq. (\ref{eq:is-1}). Due to the rapidly oscillating terms in the convolution,
the second and third terms in the above equations contribute negligibly to the \( Z \)-gate operation.
Therefore, the effective dynamics can be described by
\[
\dot{\vec{C}}(t) = -i\,G_{z}\bar{\sigma}_{z} \otimes \bar{I}_{2}\vec{C}(t),
\]
 which represents a perfect \( Z \)-gate operation on the IPQs without decoherence effects.

\subsubsection{The X-Gate operation.}

We consider the \( X \)-gate operation with \( G_{z} = 0 \). Since the information needs to be
transferred between \(  {c}_{0}^{0} \) and \(  {c}_{1}^{0} \), we must account for all the
expansion coefficients of \(  {c}_{0}^{0} \) and \(  {c}_{1}^{0} \), whose dynamics are
described by Eqs. (\ref{eq:c00-1} - \ref{eq:c11-1}), i.e.,
\begin{eqnarray*}
\dot{C}_{0}^{0}(t) & = & -i\,\mu(t)C_{0}^{0}(t) - i\,G_{x}C_{0}^{1}(t) - \int_{0}^{t}\text{d}s\,f_{0}\left(t-s\right)\left(\bar{C}_{0}^{0\,*}(s) + C_{0}^{0}(s)\right), \\
\dot{C}_{1}^{0}(t) & = & -i\,\mu(t)C_{1}^{0}(t) - i\,G_{x}C_{1}^{1}(t) - \int_{0}^{t}\text{d}s\,f_{0}\left(t-s\right)\left(\bar{C}_{1}^{0\,*}(s) + C_{1}^{0}(s)\right), \\
\dot{C}_{0}^{1}(t) & = & -i\,\mu(t)C_{0}^{1}(t) - i\,G_{x}C_{0}^{0}(t) - \int_{0}^{t}\text{d}s\,f_{1}\left(t-s\right)\left(\bar{C}_{0}^{1\,*}(s) + C_{0}^{1}(s)\right), \\
\dot{C}_{1}^{1}(t) & = & -i\,\mu(t)C_{1}^{1}(t) - i\,G_{x}C_{1}^{0}(t) - \int_{0}^{t}\text{d}s\,f_{1}\left(t-s\right)\left(\bar{C}_{1}^{1\,*}(s) + C_{1}^{1}(s)\right).
\end{eqnarray*}
We can rewrite these equations in vector form
\[
i\vec{C}(t) = \left(\mu(t)\bar{I}_{4} + G_{x}\bar{\sigma}_{x}\otimes\bar{I_{2}}\right)\vec{C}(t) + 2\int_{0}^{t}\text{d}s\,\bar{f}\left(t-s\right)\vec{C}(s) + \vec{c}(t).
\]
Using the unitary matrix
\[
\bar{U}_{X} = \frac{\sqrt{2}}{2}\left(\begin{array}{cccc}
1 & 0 & -1 & 0 \\
0 & 1 & 0 & -1 \\
1 & 0 & 1 & 0 \\
0 & 1 & 0 & 1
\end{array}\right),
\]
the vector equation can be transformed into
\[
\dot{\vec{D}}(t) = -i\,\left(\mu(t)\bar{I}_{4} + G_{x}\bar{\sigma}_{z}\otimes\bar{I_{2}}\right)\vec{D}(t) - \int_{0}^{t}\text{d}s\,\bar{F}\left(t-s\right)\vec{D}(s) + \vec{d}(t),
\]
with \( \vec{D}(t) = \bar{U}_{X}\vec{C} \), \( \vec{d} = \bar{U}_{X}\vec{c} \), and
\[
\bar{F} = \frac{1}{2}\left(\begin{array}{cccc}
f_{0} + f_{1} & 0 & f_{1} - f_{0} & 0 \\
0 & f_{0} + f_{1} & 0 & f_{1} - f_{0} \\
f_{1} - f_{0} & 0 & f_{0} + f_{1} & 0 \\
0 & f_{1} - f_{0} & 0 & f_{0} + f_{1}
\end{array}\right).
\]
By defining \( \tilde{\vec{D}}(t) = \vec{D}(t) \exp(-i\,\int_{0}^{t}\text{d}\tau\,\mu(\tau)) \), we arrive at
\begin{eqnarray*}
\dot{\tilde{\vec{D}}}(t) & = & -i\,G_{x}\bar{\sigma}_{z}\otimes\bar{I_{2}}\tilde{\vec{D}}(t) \\
 &  & - \exp\left(-i\,\int_{0}^{t}\text{d}\tau\,\mu(\tau)\right)\int_{0}^{t}\text{d}s\,\exp\left(i\,\int_{0}^{s}\text{d}\tau\,\mu(\tau)\right)\bar{F}\left(t-s\right)\tilde{\vec{D}}(s) + \tilde{\vec{d}}(t),
\end{eqnarray*}
where
\begin{eqnarray*}
\tilde{\vec{d}}(t) & = & -\exp\left(-i\,\int_{0}^{t}\text{d}\tau\,\mu(\tau)\right)\int_{0}^{t}\text{d}s\,\exp\left(-i\,\int_{0}^{s}\text{d}\tau\,\mu(\tau)\right)\bar{F}\left(t-s\right)\vec{d}(s).
\end{eqnarray*}
With the slow variables defined as
\begin{equation}
\tilde{\bar{C}}_{i}^{j\,*}(s) = \bar{C}_{i}^{j\,*}(s)\exp\!\left(i\,\int_{0}^{s}\text{d}\tau\,\mu(\tau)\right),
\label{eq:ci0b}
\end{equation}
the rapid oscillations caused by \(\exp\!\left(\pm i\,\int_{0}^{s}\text{d}\tau\,\mu(\tau)\right)\)
result in the cancellation of the last two terms in \(\dot{\tilde{\vec{D}}}(t)\) during the evolution.
 Then, we obtain a more concise effective form.
\[
\dot{\vec{D}}(t) = -i\,G_{x}\bar{\sigma}_{z}\otimes\bar{I_{2}}\vec{D}(t).
\]
When we transform the variable \( \vec{D} \) back into \( \vec{C} = \bar{U}_{D}^{\dagger}\vec{D} \), we arrive at a perfect \( X \)-gate operation:
\[
\dot{\vec{C}}(t) = -i\,G_{x}\bar{\sigma}_{x}\otimes\bar{I_{2}}\vec{C}(t).
\]

\subsection{Gate-operation dynamics}
In this subsection, we consider a simplified case where the counter-rotating wave terms
in the interaction Hamiltonian are neglected. At this time, the annihilation operators
can be expanded solely in terms of the initial annihilation operators, i.e.,
\begin{eqnarray*}
 {c}_{j}(t) & = & \sum_{i=0,1}C_{i}^{j}(t) {c}_{i}^{0}+\sum_{i,\alpha}B_{i,\alpha}^{j}(t) {b}_{i,\alpha}^{0},\\
 {c}_{j}^{\dagger}(t) & = & \sum_{i=0,1}C_{i}^{j*}(t) {c}_{i}^{0\,\dagger}+\sum_{i,\alpha}B_{i,\alpha}^{j*}(t) {b}_{i,\alpha}^{0\,\dagger}.
\end{eqnarray*}
The dynamical equations of the coefficients reads
\begin{eqnarray}
\dot{C}_{i}^{0}(t) & = & -i\,\left(\mu(t)-G_{z}\right)C_{i}^{0}(t)-i\,G_{x}C_{i}^{1}(t)-\int\text{d}s\,f_{0}\left(t-s\right)C_{i}^{0}(s),\nonumber \\
\dot{B}_{0,\alpha}^{0}(t) & = & -i\,\left(\mu(t)-G_{z}\right)B_{0,\alpha}^{0}(t)-i\,G_{x}B_{0,\alpha}^{1}(t)-\int_{0}^{t}\text{d}s\,f_{0}\left(t-s\right)B_{0,\alpha}^{0}(s)\nonumber \\
 &  & -i\,g_{0,\alpha}\exp\left(-i\:\omega_{0,\alpha}t\right),\label{eq:dbi0B-1}\\
\dot{B}_{1,\alpha}^{0}(t) & = & -i\,\left(\mu(t)-G_{z}\right)B_{1,\alpha}^{0}(t)-i\,G_{x}B_{1,\alpha}^{1}(t)-\int_{0}^{t}\text{d}s\,f_{0}\left(t-s\right)B_{1,\alpha}^{0}(s)\nonumber
\end{eqnarray}
\begin{eqnarray}
\dot{C}_{i}^{1}(t) & = & -i\,\left(\mu(t)+G_{z}\right)C_{i}^{1}(t)-i\,G_{x}C_{i}^{0}(t)-\int_{0}^{t}\text{d}s\,f_{1}\left(t-s\right)C_{i}^{1}(s),\nonumber \\
\dot{B}_{0,\alpha}^{1}(t) & = & -i\,\left(\mu(t)+G_{z}\right)B_{0,\alpha}^{1}(t)-i\,G_{x}B_{0,\alpha}^{0}(t)-\int_{0}^{t}\text{d}s\,f_{1}\left(t-s\right)B_{0,\alpha}^{1}(s),\nonumber \\
\dot{B}_{1,\alpha}^{1}(t) & = & -i\,\left(\mu(t)+G_{z}\right)B_{1,\alpha}^{1}(t)-i\,G_{x}B_{1,\alpha}^{0}(t)-\int_{0}^{t}\text{d}s\,f_{1}\left(t-s\right)B_{1,\alpha}^{1}(s),\nonumber \\
 &  & -i\,g_{1,\alpha}\exp\left(-i\:\omega_{1,\alpha}t\right).\label{eq:dbi1-1}
\end{eqnarray}
In what follows, we simulate the average value of the observable variants
$\langle J_{x}(t)\rangle$, $\langle J_{y}(t)\rangle$, and $\langle J_{z}(t)\rangle$,
where $\langle J_{k}(t)\rangle=\text{Tr}\{J_{k}(t)\rho_{\text{tot}}\}$.
We set that the system and heat reserviors are not correlated eachother
and the total density matrix is written as $\rho_{\text{tot}}=\rho_{\text{S}}(0)\otimes\rho_{\text{B}}^{0}(0)\otimes\rho_{\text{B}}^{1}(0)$,
and the reservoir is prepared on the thermal equilibrium state
\[
\rho_{B}^{k}(0)=\sum_{n,\alpha}\frac{e^{-\beta_{k}\omega_{k,\alpha}}}{1-e^{-\beta_{k}\omega_{k,\alpha}}}\left|n_{k,\alpha}\right\rangle \left\langle n_{k,\alpha}\right|,
\]
where $\beta_{k}$ is the inverse reservoir temperature and $\left|n_{k}\right\rangle $
is the fock state with particle number $n_{k}$.

\subsubsection{The Z-gate operation.}

We analyze the expected values of the Pauli operators within the context of the Z-gate operation.
The z-component is defined as
\[
\left\langle J_{z}(t) \right\rangle = \left\langle  {c}_{1}^{\dagger}(t)  {c}_{1}(t) -  {c}_{0}^{\dagger}(t)  {c}_{0}(t) \right\rangle,
\]
which can be expressed as
\[
\begin{aligned}
&= \sum_{i,j=0}^{1} \left( C_{j}^{1*}(t) C_{i}^{1}(t) - C_{j}^{0*}(t) C_{i}^{0}(t) \right) \left\langle  {c}_{j}^{0\dagger}  {c}_{i}^{0} \right\rangle_{\text{S}} \\
& \quad + \sum_{i,\alpha} \left( |B_{i,\alpha}^{1}(t)|^{2} - |B_{i,\alpha}^{0}(t)|^{2} \right) \left\langle  {b}_{i,\alpha}^{0\dagger}  {b}_{i,\alpha}^{0} \right\rangle_{\text{B}},
\end{aligned}
\]
where we have considered that \(\text{Tr}_{B} \{  {b}_{j,\beta}^{0\dagger}
 {b}_{i,\alpha}^{0} \rho_{\text{B}}(0) \} = \delta_{\alpha\beta} \delta_{ij}\).
 \(B_{j,\beta}^{i}\) can be calculated using the Laplace
transformation. By applying the Laplace transformation to
Eqs. (\ref{eq:dbi0B-1}) and (\ref{eq:dbi1-1}), we obtain
\[
\begin{aligned}
\tilde{B}_{1,\alpha}^{1}(t_{1}) & = \frac{-i g_{1,\alpha}}{\left(t_{1} + i \omega_{1,\alpha}\right)\left(t_{1} + \tilde{f}_{1}(t_{1}) + i(\mu + G_{z})\right)}, \\
\tilde{B}_{0,\alpha}^{0}(t_{1}) & = \frac{-i g_{0,\alpha}}{\left(t_{1} + i \omega_{0,\alpha}\right)\left(t_{1} + \tilde{f}_{0}(t_{1}) + i(\mu - G_{z})\right)},
\end{aligned}
\]
where \(\tilde{f}_{j}(t_{1})\) represents the Laplace transformation of the correlation
 function for the \(j\)-th reservoir, and we have assumed \(B_{i,\alpha}^{j}(0) = 0\).
 Since \(B_{i,\alpha}^{j}(0) = 0\), it follows that \(B_{1}^{0}(t) = B_{0}^{1}(t) = 0\).
 If the correlation function \(f_{j}(t-s)\) is specified, the solution can be derived by
 taking the inverse Laplace transformation of \(\tilde{B}_{i,\alpha}^{j}(t_{1})\),
  i.e., \(B_{j,\beta}^{i}(t) = \mathcal{L}^{-1}[\tilde{B}_{i,\alpha}^{j}(t_{1})]\).
By introducing
\[
\begin{aligned}
\tilde{B'}_{1,\alpha}^{1}(t_{1}) & = \frac{1}{\left(t_{1} + i \omega_{1,\alpha}\right)\left(t_{1} + \tilde{f}_{1}(t_{1}) + i(\mu + G_{z})\right)}, \\
\tilde{B'}_{0,\alpha}^{0}(t_{1}) & = \frac{1}{\left(t_{1} + i \omega_{0,\alpha}\right)\left(t_{1} + \tilde{f}_{0}(t_{1}) + i(\mu - G_{z})\right)},
\end{aligned}
\]
we see that \(B_{j,\beta}^{i}(t) = -i g_{i,\alpha} B'{}_{j,\beta}^{i}(t)\) with
\(B'{}_{j,\beta}^{i}(t) = \mathcal{L}^{-1}[\tilde{B'}_{j,\alpha}^{i}(t_{1})]\),
leading to the expressions reported in the main text, Eqs.~(\ref{Jzind}) and (\ref{Jxind}).

where \(B_{1}^{0}(t) = B_{1}^{0}(t) = 0\) and \(\left\langle  {b}_{j,\beta}^{0\dagger}
 {b}_{i,\alpha}^{0} \right\rangle_{\text{B}} = \delta_{\alpha\beta} \delta_{ij}
\frac{1}{e^{\beta_{0} \omega_{j,\alpha}} - 1}\) have been utilized.

\subsubsection{The X-gate operation.}

In this case, the dynamics of $B_{j,\alpha}^{i}(t)$
is different to the Z-gate operation case. We divid the dynamical
equations into two independent groups. One is
\begin{eqnarray*}
\dot{B}_{0,\alpha}^{0}(t) & = & -i\,\mu B_{0,\alpha}^{0}(t)-i\,G_{x}B_{0,\alpha}^{1}(t)-\int_{0}^{t}\text{d}s\,f_{0}\left(t-s\right)B_{0,\alpha}^{0}(s)\\
 &  & -i\,g_{0,\alpha}\exp\left(-i\:\omega_{0,\alpha}t\right),\\
\dot{B}_{0,\alpha}^{1}(t) & = & -i\,\mu B_{0,\alpha}^{1}(t)-i\,G_{x}B_{0,\alpha}^{0}(t)-\int_{0}^{t}\text{d}s\,f_{1}\left(t-s\right)B_{0,\alpha}^{1}(s),
\end{eqnarray*}
and another one reads
\begin{eqnarray*}
\dot{B}_{1,\alpha}^{1}(t) & = & -i\,\mu B_{1,\alpha}^{1}(t)-i\,G_{x}B_{1,\alpha}^{0}(t)-\int_{0}^{t}\text{d}s\,f_{1}\left(t-s\right)B_{1,\alpha}^{1}(s),\\
 &  & -i\,g_{1,\alpha}\exp\left(-i\:\omega_{1,\alpha}t\right).\\
\dot{B}_{1,\alpha}^{0}(t) & = & -i\,\mu B_{1,\alpha}^{0}(t)-i\,G_{x}B_{1,\alpha}^{1}(t)-\int_{0}^{t}\text{d}s\,f_{0}\left(t-s\right)B_{1,\alpha}^{0}(s),
\end{eqnarray*}
which can be solved by the Laplace transformation.
\begin{eqnarray*}
\tilde{B}_{0,\alpha}^{0}(t_{1}) & = & \frac{i\,g_{0,\alpha}\,\left(t_{1}+i\,\mu+\tilde{f}_{1}(t_{1})\right)}{\left(t_{1}+i\,\omega_{0,\alpha}\right)\,\left(G_{x}^{2}+\left(t_{1}+i\,\mu+\tilde{f}_{0}(t_{1})\right)\,\left(t_{1}+i\,\mu+\tilde{f}_{1}(t_{1})\right)\right)},\\
\tilde{B}_{0,\alpha}^{1}(t_{1}) & = & \frac{i\,G_{x}\,g_{0,\alpha}}{\left(t_{1}+i\,\omega_{0,\alpha}\right)\,\left(G_{x}^{2}+\left(t_{1}+i\,\mu+\tilde{f}_{0}(t_{1})\right)\,\left(t_{1}+i\,\mu+\tilde{f}_{1}(t_{1})\right)\right)},\\
\tilde{B}_{1,\alpha}^{0}(t_{1}) & = & \frac{i\,G_{x}\,g_{1,\alpha}}{\left(t_{1}+i\,\omega_{1,\alpha}\right)\,\left(G_{x}^{2}+\left(t_{1}+i\,\mu+\tilde{f}_{0}(t_{1})\right)\,\left(t_{1}+i\,\mu+\tilde{f}_{1}(t_{1})\right)\right)},\\
\tilde{B}_{1\alpha}^{1}(t_{1}) & = & \frac{i\,g_{1,\alpha}\,\left(t_{1}+i\,\mu+\tilde{f}_{0}(t_{1})\right)}{\left(t_{1}+i\,\omega_{1,\alpha}\right)\,\left(G_{x}^{2}+\left(t_{1}+i\,\mu+\tilde{f}_{0}(t_{1})\right)\,\left(t_{1}+i\,\mu+\tilde{f}_{1}(t_{1})\right)\right)}.
\end{eqnarray*}
where \(\tilde{f}_{k}(t_{1}) = \mathcal{L}[f_{k}(t)]\) is the Laplace transform of \(f_{k}(t)\),
and the initial conditions\(B_{j}^{i}(0) = 0\) have been used. If \(f_{k}(t)\) is known,
the solutions can be obtained via the inverse Laplace transformation.
The mean values of \(\{J_{z}\}\) are given by
\begin{eqnarray*}
\left\langle J_{x}(t)\right\rangle  & = & \sum_{i,j=0,1}\left(C_{i}^{1*}(t)C_{j}^{0}(t)+C_{i}^{0*}(t)C_{j}^{1}(t)\right)\left\langle  {c}_{i}^{0\,\dagger} {c}_{j}^{0}\right\rangle _{\text{S}}\\
 & + & \sum_{i,\alpha}|g_{i,\alpha}|^{2}\left(B'{}_{i,\alpha}^{0*}(t)B_{i,\alpha}^{'1}(t)+B'{}_{i,\alpha}^{1*}(t)B'{}_{i,\alpha}^{0}(t)\right)\frac{1}{e^{\beta_{i}\omega_{i,\alpha}}-1},\\
\left\langle J_{y}(t)\right\rangle  & = & \sum_{i,j=0,1}i\,\left(C_{i}^{1*}(t)C_{j}^{0}(t)-C_{i}^{0*}(t)C_{j}^{1}(t)\right)\left\langle  {c}_{i}^{0\,\dagger} {c}_{j}^{0}\right\rangle _{\text{S}}\\
 & + & \sum_{i,\alpha}i\,|g_{i,\alpha}|^{2}\left(B'{}_{i,\alpha}^{0*}(t)B'{}_{i,\alpha}^{1}(t)-B'{}_{i,\alpha}^{1*}(t)B'{}_{i,\alpha}^{0}(t)\right)\frac{1}{e^{\beta_{i}\omega_{i,\alpha}}-1},\\
\left\langle J_{z}(t)\right\rangle  & = & \sum_{i,j=0,1}\left(C_{j}^{1*}(t)C_{i}^{1}(t)-C_{j}^{0*}(t)C_{i}^{0}(t)\right)\left\langle  {c}_{j}^{0\dagger} {c}_{i}^{0}\right\rangle _{\text{S}}\\
 &  & +\sum_{i,\alpha}|g_{i,\alpha}|^{2}\left(B'{}_{i,\alpha}^{1*}(t)B'{}_{i,\alpha}^{1}(t)-B'{}_{i,\alpha}^{0*}(t)B'{}_{i,\alpha}^{0}(t)\right)\frac{1}{e^{\beta_{i}\omega_{i,\alpha}}-1},
\end{eqnarray*}
with $\left\langle  {b}_{j,\beta}^{0\dagger} {b}_{i,\alpha}^{0}\right\rangle _{\text{B}}=\delta_{\alpha\beta}\delta_{ij}\frac{1}{e^{\beta_{j}\omega_{j,\alpha}}-1}$.

\section{Initial states and expectation values}\label{app:initial}
We assume that the initial system state is a pure state which can
be written as $\rho_{\text{S}}(0)=|\psi(0)\rangle\langle\psi(0)|$
with $|\psi(0)\rangle=\alpha_{0} {c}_{0}^{0\dagger}|00\rangle+\alpha_{1} {c}_{1}^{0\dagger}|00\rangle$.
Here, $\alpha_{0}$ and $\alpha_{1}$ are arbitrary complex amplitudes
fulfilling $|\alpha_1|^{2}+|\alpha_0|^{2}=1$. Thus we have
\begin{eqnarray*}
\left\langle  {a}_{i}^{0\,\dagger} {a}_{j}^{0}\right\rangle _{\text{S}} & = & \text{Tr}_{\text{S}}\left\{  {a}_{i}^{0\,\dagger} {a}_{j}^{0}\rho_{\text{S}}(0)\right\} \\
 & = & \langle\psi(0)| {a}_{i}^{0\,\dagger} {a}_{j}^{0}|\psi(0)\rangle\\
 & = & \frac{1}{2}\sum_{k,l}\alpha_{k}^{*}\alpha_{l}\langle00|\left( {c}_{k}^{0} {c}_{1}^{0\dagger} {c}_{1}^{0} {c}_{l}^{0\dagger}-(-1)^{i} {c}_{k}^{0} {c}_{0}^{0\dagger} {c}_{1}^{0} {c}_{l}^{0\dagger}-(-1)^{j} {c}_{k}^{0} {c}_{1}^{0\dagger} {c}_{0}^{0} {c}_{l}^{0\dagger}+(-1)^{i+j} {c}_{k}^{0} {c}_{0}^{0\dagger} {c}_{0}^{0} {c}_{l}^{0\dagger}\right) {c}_{l}^{0\dagger}|00\rangle\\
 & = & \frac{1}{2}\sum_{k,l}\alpha_{k}^{*}\alpha_{l}\left(\delta_{k1}\delta_{l1}-(-1)^{i}\delta_{k0}\delta_{l1}-(-1)^{j}\delta_{k1}\delta_{l0}+(-1)^{i+j}\delta_{k0}\delta_{l0}\right)\\
 & = & \frac{1}{2}\left(\alpha_{1}^{*}\alpha_{1}-(-1)^{i}\alpha_{0}^{*}\alpha_{1}-(-1)^{j}\alpha_{1}^{*}\alpha_{0}+(-1)^{i+j}\alpha_{0}^{*}\alpha_{0}\right),
\end{eqnarray*}
which can be used to determine the mean values of the system part.
The mixed initial system state can be written as $\rho_{\text{S}}(0)=\sum_{kl}\rho_{kl} {c}_{k}^{0\dagger}|00\rangle\langle00| {c}_{l}^{0}$,
which leads to
\begin{eqnarray*}
\left\langle  {a}_{i}^{0\,\dagger} {a}_{j}^{0}\right\rangle _{\text{S}} & = & \text{Tr}_{\text{S}}\left\{  {a}_{i}^{0\,\dagger} {a}_{j}^{0}\rho_{\text{S}}(0)\right\} \\
 & = & \frac{1}{2}\sum_{kl}\rho_{kl}\langle00| {c}_{l}^{0}\left( {c}_{1}^{\dagger}-(-1)^{i} {c}_{0}^{\dagger}\right)\left( {c}_{1}-(-1)^{j} {c}_{0}\right) {c}_{k}^{0\dagger}|00\rangle\\
 & = & \frac{1}{2}\sum_{kl}\rho_{kl}\langle00| {c}_{l}^{0}\left( {c}_{1}^{\dagger} {c}_{1}-(-1)^{j} {c}_{1}^{\dagger} {c}_{0}-(-1)^{i} {c}_{0}^{\dagger} {c}_{1}+(-1)^{j+i} {c}_{0}^{\dagger} {c}_{0}\right) {c}_{k}^{0\dagger}|00\rangle\\
 & = & \frac{1}{2}\left(\rho_{11}-(-1)^{j}\rho_{01}-(-1)^{i}\rho_{10}+(-1)^{j+i}\rho_{00}\right)
\end{eqnarray*}

\twocolumngrid


\begin{thebibliography}{0}%
\makeatletter
\providecommand \@ifxundefined [1]{%
 \@ifx{#1\undefined}
}%
\providecommand \@ifnum [1]{%
 \ifnum #1\expandafter \@firstoftwo
 \else \expandafter \@secondoftwo
 \fi
}%
\providecommand \@ifx [1]{%
 \ifx #1\expandafter \@firstoftwo
 \else \expandafter \@secondoftwo
 \fi
}%
\providecommand \natexlab [1]{#1}%
\providecommand \enquote  [1]{``#1''}%
\providecommand \bibnamefont  [1]{#1}%
\providecommand \bibfnamefont [1]{#1}%
\providecommand \citenamefont [1]{#1}%
\providecommand \href@noop [0]{\@secondoftwo}%
\providecommand \href [0]{\begingroup \@sanitize@url \@href}%
\providecommand \@href[1]{\@@startlink{#1}\@@href}%
\providecommand \@@href[1]{\endgroup#1\@@endlink}%
\providecommand \@sanitize@url [0]{\catcode `\\12\catcode `\$12\catcode
  `\&12\catcode `\#12\catcode `\^12\catcode `\_12\catcode `\%12\relax}%
\providecommand \@@startlink[1]{}%
\providecommand \@@endlink[0]{}%
\providecommand \url  [0]{\begingroup\@sanitize@url \@url }%
\providecommand \@url [1]{\endgroup\@href {#1}{\urlprefix }}%
\providecommand \urlprefix  [0]{URL }%
\providecommand \Eprint [0]{\href }%
\providecommand \doibase [0]{http://dx.doi.org/}%
\providecommand \selectlanguage [0]{\@gobble}%
\providecommand \bibinfo  [0]{\@secondoftwo}%
\providecommand \bibfield  [0]{\@secondoftwo}%
\providecommand \translation [1]{[#1]}%
\providecommand \BibitemOpen [0]{}%
\providecommand \bibitemStop [0]{}%
\providecommand \bibitemNoStop [0]{.\EOS\space}%
\providecommand \EOS [0]{\spacefactor3000\relax}%
\providecommand \BibitemShut  [1]{\csname bibitem#1\endcsname}%
\let\auto@bib@innerbib\@empty
\end{thebibliography}%


\begin{thebibliography}{99}

\bibitem{Schlosshauer2019}
M. Schlosshauer,
Quantum decoherence,
Phys. Rep. \textbf{831}, 1 (2019).

\bibitem{Ripoll2003}
J. J. Garc\'{\i}a-Ripoll and J. I. Cirac,
Quantum computation with cold bosonic atoms in an optical lattice,
Philos. Trans. R. Soc. A \textbf{361}, 1537 (2003).

\bibitem{Cao2023}
S. Cao \textit{et al.},
Generation of genuine entanglement up to 51 superconducting qubits,
Nature (London) \textbf{619}, 738 (2023).

\bibitem{Arunkumar2023}
N. Arunkumar, K. S. Olsson, J. T. Oon, C. A. Hart, D. B. Bucher,
D. R. Glenn, M. D. Lukin, H. Park, D. Ham, and R. L. Walsworth,
Quantum logic enhanced sensing in solid-state spin ensembles,
Phys. Rev. Lett. \textbf{131}, 100801 (2023).

\bibitem{Jain2024}
S. Jain \textit{et al.},
Penning micro-trap for quantum computing,
Nature (London) \textbf{627}, 510 (2024).

\bibitem{Leon2021}
N. P. de Leon, K. M. Itoh, D. Kim, K. K. Mehta, T. E. Northup,
H. Paik, B. S. Palmer, N. Samarth, S. Sangtawesin, and D. W. Steuerman,
Materials challenges and opportunities for quantum computing hardware,
Science \textbf{372}, eabb2823 (2021).

\bibitem{PRX}
M. H. Michael, M. Silveri, R. T. Brierley, V. V. Albert,
J. Salmilehto, L. Jiang, and S. M. Girvin,
New class of quantum error-correcting codes for a bosonic mode,
Phys. Rev. X \textbf{6}, 031006 (2016).

\bibitem{Paetznick2021}
A. Paetznick \textit{et al.},
Demonstration of logical qubits and repeated error correction with
better-than-physical error rates,
arXiv:2404.02280 (2024).

\bibitem{Preskill2025}
H. Putterman \textit{et al.},
Hardware-efficient quantum error correction via concatenated bosonic qubits,
Nature (London) \textbf{638}, 927 (2025).

\bibitem{Binormial2023}
Z. Ni \textit{et al.},
Beating the break-even point with a discrete-variable-encoded logical qubit,
Nature (London) \textbf{616}, 56 (2023).

\bibitem{Brock2025}
B. L. Brock, S. Singh, A. Eickbusch, V. V. Sivak, A. Z. Ding,
L. Frunzio, S. M. Girvin, and M. H. Devoret,
Quantum error correction of qudits beyond break-even,
Nature (London) \textbf{641}, 612 (2025).

\bibitem{Teoh2023}
J. D. Teoh \textit{et al.},
Dual-rail encoding with superconducting cavities,
Proc. Natl. Acad. Sci. USA \textbf{120}, e2221736120 (2023).

\bibitem{Chou2024}
K. S. Chou \textit{et al.},
A superconducting dual-rail cavity qubit with erasure-detected logical
measurements,
Nat. Phys. \textbf{20}, 1454 (2024).

\bibitem{Grassl1997}
M. Grassl, T. Beth, and T. Pellizzari,
Codes for the quantum erasure channel,
Phys. Rev. A \textbf{56}, 33 (1997).

\bibitem{Knill2001}
E. Knill, R. Laflamme, and G. J. Milburn,
A scheme for efficient quantum computation with linear optics,
Nature (London) \textbf{409}, 46 (2001).

\bibitem{Shapiro2006}
J. H. Shapiro,
Single-photon Kerr nonlinearities do not help quantum computation,
Phys. Rev. A \textbf{73}, 062305 (2006).

\bibitem{Li2015}
Y. Li, P. C. Humphreys, G. J. Mendoza, and S. C. Benjamin,
Resource costs for fault-tolerant linear optical quantum computing,
Phys. Rev. X \textbf{5}, 041007 (2015).

\bibitem{Bartolucci2023}
S. Bartolucci \textit{et al.},
Fusion-based quantum computation,
Nat. Commun. \textbf{14}, 912 (2023).

\bibitem{Berdou2023}
C. Berdou \textit{et al.},
One hundred second bit-flip time in a two-photon dissipative oscillator,
PRX Quantum \textbf{4}, 020350 (2023).

\bibitem{Reinhold2020}
P. Reinhold \textit{et al.},
Error-corrected gates on an encoded qubit,
Nat. Phys. \textbf{16}, 822 (2020).

\bibitem{Adam2014}
M. Bartkowiak, L.-A. Wu, and A. Miranowicz,
Quantum circuits for amplification of Kerr nonlinearity via quadrature
squeezing,
J. Phys. B: At. Mol. Opt. Phys. \textbf{47}, 145501 (2014).

\bibitem{Wu2009}
L.-A. Wu, A. Miranowicz, X. B. Wang, Y.-x. Liu, and F. Nori,
Perfect function transfer and interference effects in interacting boson
lattices,
Phys. Rev. A \textbf{80}, 012332 (2009).

\bibitem{Lin2020}
Y. Lin, D. R. Leibrandt, D. Leibfried, and C. W. Chou,
Quantum entanglement between an atom and a molecule,
Nature (London) \textbf{581}, 273 (2020).

\bibitem{Sun2014}
L. Sun \textit{et al.},
Tracking photon jumps with repeated quantum non-demolition parity
measurements,
Nature (London) \textbf{511}, 444 (2014).

\bibitem{Proctor2017}
T. Proctor, K. Rudinger, K. Young, M. Sarovar, and R. Blume-Kohout,
What randomized benchmarking actually measures,
Phys. Rev. Lett. \textbf{119}, 130502 (2017).

\bibitem{Friesen2017}
M. Friesen, J. Ghosh, M. A. Eriksson, and S. N. Coppersmith,
A decoherence-free subspace in a charge quadrupole qubit,
Nat. Commun. \textbf{8}, 15923 (2017).

\bibitem{Wu2002}
L.-A. Wu, M. S. Byrd, and D. A. Lidar,
Efficient universal leakage elimination for physical and encoded qubits,
Phys. Rev. Lett. \textbf{89}, 127901 (2002).

\bibitem{Jing2015}
J. Jing, L.-A. Wu, M. S. Byrd, J. Q. You, T. Yu, and Z. M. Wang,
Nonperturbative leakage elimination operators and control of a three-level
system,
Phys. Rev. Lett. \textbf{114}, 190502 (2015).

\bibitem{Ezzell2023}
N. Ezzell, B. Pokharel, L. Tewala, G. Quiroz, and D. A. Lidar,
Dynamical decoupling for superconducting qubits: A performance survey,
Phys. Rev. Appl. \textbf{20}, 064027 (2023).

\bibitem{Tosta2019}
A. D. C. Tosta, D. J. Brod, and E. F. Galv\~{a}o,
Quantum computation from fermionic anyons on a one-dimensional lattice,
Phys. Rev. A \textbf{99}, 062335 (2019).

\bibitem{WuLidar02}
L.-A. Wu and D. A. Lidar,
Qubits as parafermions,
J. Math. Phys. \textbf{43}, 4506 (2002).

\bibitem{Tsunoda2023}
T. Tsunoda, J. D. Teoh, W. D. Kalfus, S. J. de Graaf,
B. J. Chapman, J. C. Curtis, N. Thakur, S. M. Girvin, and R. J. Schoelkopf,
Error-detectable bosonic entangling gates with a noisy ancilla,
PRX Quantum \textbf{4}, 020354 (2023).

\bibitem{Kubica2023}
A. Kubica, A. Haim, Y. Vaknin, F. Brand\~{a}o, and A. Retzker,
Erasure qubits: Overcoming the $T_1$ limit in superconducting circuits,
Phys. Rev. X \textbf{13}, 041022 (2023).

\bibitem{Lukens2017}
J. M. Lukens and P. Lougovski,
Frequency-encoded photonic qubits for scalable quantum information
processing,
Optica \textbf{4}, 8 (2017).

\bibitem{Ilves2020}
J. Ilves, S. Kono, Y. Sunada, S. Yamazaki, M. Kim, K. Koshino,
and Y. Nakamura,
On-demand generation and characterization of a microwave time-bin qubit,
npj Quantum Inf. \textbf{6}, 34 (2020).

\bibitem{Scala2024}
F. Scala, D. Nigro, and D. Gerace,
Deterministic entangling gates with nonlinear quantum photonic
interferometers,
Commun. Phys. \textbf{7}, 118 (2024).

\bibitem{Shor1995}
P. W. Shor,
Scheme for reducing decoherence in quantum computer memory,
Phys. Rev. A \textbf{52}, R2493 (1995).

\bibitem{Anderson2021}
C. Ryan-Anderson \textit{et al.},
Realization of real-time fault-tolerant quantum error correction,
Phys. Rev. X \textbf{11}, 041058 (2021).

\bibitem{Neeve2022}
B. de Neeve, T. L. Nguyen, T. Behrle, and J. P. Home,
Error correction of a logical grid state qubit by dissipative pumping,
Nat. Phys. \textbf{18}, 296 (2022).

\bibitem{Google2024}
Google Quantum AI and Collaborators,
Quantum error correction below the surface code threshold,
Nature (London) \textbf{638}, 920 (2025).

\bibitem{WuL02}
L.-A. Wu and D. A. Lidar,
Creating decoherence-free subspaces using strong and fast pulses,
Phys. Rev. Lett. \textbf{88}, 207902 (2002).

\bibitem{Terhal2015}
B. M. Terhal,
Quantum error correction for quantum memories,
Rev. Mod. Phys. \textbf{87}, 307 (2015).

\bibitem{Magesan2011}
E. Magesan, J. M. Gambetta, and J. Emerson,
Scalable and robust randomized benchmarking of quantum processes,
Phys. Rev. Lett. \textbf{106}, 180504 (2011).

\bibitem{WuLi1}
Y. Li, L.-A. Wu, Y.-D. Wang, and L.-P. Yang,
Nondeterministic ultrafast ground-state cooling of a mechanical resonator,
Phys. Rev. B \textbf{84}, 094502 (2011).

\bibitem{WuLi2}
J.-S. Xu, M.-H. Yung, X.-Y. Xu, S. Boixo, Z.-W. Zhou, C.-F. Li,
A. Aspuru-Guzik, and G.-C. Guo,
Demon-like algorithmic quantum cooling and its realization with quantum
optics,
Nat. Photonics \textbf{8}, 113 (2014).

\bibitem{Egan2021}
L. Egan, D. M. Debroy, C. Noel, A. Risinger, D. Zhu, D. Biswas,
M. Newman, M. Li, K. R. Brown, M. Cetina, and C. Monroe,
Fault-tolerant operation of a quantum error-correction code,
Nature (London) \textbf{598}, 281 (2021).

\bibitem{Yang2001}
C.-P. Yang and J. Gea-Banacloche,
Three-qubit quantum error-correction scheme for collective decoherence,
Phys. Rev. A \textbf{63}, 022311 (2001).

\bibitem{Otterpohl2022}
F. Otterpohl, P. Nalbach, and M. Thorwart,
The hidden phase of the spin-boson model,
Phys. Rev. Lett. \textbf{129}, 120406 (2022).

\bibitem{Kaur2021}
K. Kaur, T. S\'{e}pulcre, N. Roch, I. Snyman, S. Florens, and S. Bera,
Spin-boson quantum phase transition in multilevel superconducting qubits,
Phys. Rev. Lett. \textbf{127}, 237702 (2021).

\bibitem{Wang2024}
Z. M. Wang, S. L. Wu, M. S. Byrd, and L.-A. Wu,
Analytical solutions of bilinear bosonic models and exact characterization
of non-Markovian dynamics,
Phys. Rev. A \textbf{111}, 062206 (2025).

\bibitem{Burgarth2021}
D. Burgarth, P. Facchi, M. Ligab\`{o}, and D. Lonigro,
Hidden non-Markovianity in open quantum systems,
Phys. Rev. A \textbf{103}, 012203 (2021).

\bibitem{Suter2016}
D. Suter and G. A. \'{A}lvarez,
Colloquium: Protecting quantum information against environmental noise,
Rev. Mod. Phys. \textbf{88}, 041001 (2016).

\bibitem{Zheng2020}
S. S. Zheng, Q. Y. He, M. S. Byrd, and L.-A. Wu,
Nonperturbative leakage elimination for a logical qubit encoded in a
mechanical oscillator,
Phys. Rev. Res. \textbf{2}, 033378 (2020).

\bibitem{NP}
J. Alicea, Y. Oreg, G. Refael, F. von Oppen, and M. P. A. Fisher,
Non-Abelian statistics and topological quantum information processing in
1D wire networks,
Nat. Phys. \textbf{7}, 412 (2011).

\bibitem{Gradshteyn1980}
I. S. Gradshteyn and I. M. Ryzhik,
\textit{Table of Integrals, Series, and Products}
(Academic Press, New York, 1980).

\bibitem{Brown1993}
J. W. Brown and R. V. Churchill,
\textit{Fourier Series and Boundary Value Problems}
(McGraw-Hill, New York, 1993).

\bibitem{Heuck2020}
M. Heuck, K. Jacobs, and D. R. Englund,
Controlled-phase gate using dynamically coupled cavities and optical
nonlinearities,
Phys. Rev. Lett. \textbf{124}, 160501 (2020).

\bibitem{Bloch2008}
I. Bloch, J. Dalibard, and W. Zwerger,
Many-body physics with ultracold gases,
Rev. Mod. Phys. \textbf{80}, 885 (2008).

\bibitem{Blais2021}
A. Blais, A. L. Grimsmo, S. M. Girvin, and A. Wallraff,
Circuit quantum electrodynamics,
Rev. Mod. Phys. \textbf{93}, 025005 (2021).

\bibitem{Aspelmeyer2014}
M. Aspelmeyer, T. J. Kippenberg, and F. Marquardt,
Cavity optomechanics,
Rev. Mod. Phys. \textbf{86}, 1391 (2014).

\bibitem{Koch2007}
J. Koch, T. M. Yu, J. Gambetta, A. A. Houck, D. I. Schuster,
J. Majer, A. Blais, M. H. Devoret, S. M. Girvin, and R. J. Schoelkopf,
Charge-insensitive qubit design derived from the Cooper pair box,
Phys. Rev. A \textbf{76}, 042319 (2007).

\end{thebibliography}
\end{document}